\documentclass[conference]{IEEEtran}
\usepackage{etex}

\usepackage[noadjust]{cite}
\usepackage[utf8]{inputenc} 

\usepackage{graphicx}

\usepackage{amsmath}
\usepackage{amsthm}
\usepackage{amsfonts,amssymb}
\usepackage{array}
\usepackage{proof}

\usepackage[all]{xy}

\usepackage{url}
\usepackage{hyperref}

\usepackage{graphicx}
\usepackage{tikz-cd}
\usetikzlibrary{arrows}
\usetikzlibrary{calc}
\usetikzlibrary{trees}
\usetikzlibrary{shapes}
\usetikzlibrary{decorations.markings}
\usetikzlibrary{positioning}

\tikzset{
    fromroot/.style={draw,postaction={decorate},
        decoration={markings,mark=at position .55 with {\arrow{>}}}},
    toroot/.style={draw,postaction={decorate},
        decoration={markings,mark=at position .55 with {\arrow{<}}}},
}

\definecolor{lightblue}{RGB}{173,216,230}
\definecolor{indianred}{RGB}{205,92,92}

\tikzstyle{var}=[circle,fill=black,draw=black,scale=0.1]
\tikzstyle{lam}=[circle,fill=lightblue,draw=black,scale=0.6]
\tikzstyle{app}=[circle,fill=indianred,draw=black,scale=0.6]

\newcommand*{\imgcenter}[1]{\begingroup\setbox0=\hbox{#1}\parbox{\wd0}{\box0}\endgroup}
\newcommand\definand[1]{{\bf #1}}%

\newtheorem{theorem}{Theorem}[section]
\newtheorem{lemma}[theorem]{Lemma}
\newtheorem{corollary}[theorem]{Corollary}

\newtheorem{proposition}[theorem]{Proposition}
\newtheorem{claim}[theorem]{Claim}

\theoremstyle{definition}
\newtheorem{definition}[theorem]{Definition}

\newtheorem{example}{Example}

\newcommand\tseq[2]{#1 \longrightarrow #2}
\newcommand\fseq[2]{#1 \Longrightarrow #2}

\newcommand\Catalan[1]{C_{#1}}
\newcommand\Tam[1]{\mathsf{Y}_{#1}}
\newcommand\Int[1]{\mathcal{I}_{#1}}

\newcommand\applica[1]{A_{#1}}
\newcommand\bindtrac[1]{\Gamma_{#1}}
\newcommand\apptree[1]{\alpha[#1]}
\newcommand\binddiag[1]{\gamma[#1]}
\newcommand\bindfor[1]{\gamma^F[#1]}
\newcommand\bindtree[1]{\gamma^T[#1]}

\newcommand\atm{\mathrm{atm}}
\newcommand\foc{\mathrm{foc}}

\newcommand\set[1]{\left\{\,#1\,\right\}}

\newcommand\toF[1]{\phi[#1]}
\newcommand\actF[2]{{#1}\circledast #2}
\newcommand\deriv[1]{\mathcal{#1}}
\newcommand\toC[1]{\psi[#1]}%

\newcommand\orig{\mathrm{amb}}

\newcommand\sz[1]{|#1|}%
\newcommand\frontier[1]{\mathrm{\mathrm{fr}}(#1)}

\newcommand\irr{{\mathrm{irr}}}

\let\wild=\textvisiblespace

\begin{document}
\title{A sequent calculus for the Tamari order}

\author{\IEEEauthorblockN{Noam Zeilberger}
\IEEEauthorblockA{University of Birmingham
\\ noam.zeilberger@gmail.com
}
}

\maketitle

\begin{abstract}
We introduce a sequent calculus with a simple restriction of Lambek's product rules that precisely captures the classical Tamari order, i.e., the partial order on fully-bracketed words (equivalently, binary trees) induced by a semi-associative law (equivalently, tree rotation).
We establish a focusing property for this sequent calculus (a strengthening of cut-elimination), which yields the following coherence theorem: every valid entailment in the Tamari order has exactly one focused derivation.
One combinatorial application of this coherence theorem is a new proof of the Tutte--Chapoton formula for the number of intervals in the Tamari lattice $\Tam{n}$.
We also apply the sequent calculus and the coherence theorem to build a surprising bijection between intervals of the Tamari order and a certain fragment of lambda calculus, consisting of the $\beta$-normal planar lambda terms with no closed proper subterms.
\end{abstract}

\IEEEpeerreviewmaketitle

\section{Introduction}
\label{sec:intro}

\subsection{The Tamari order, and Tamari lattices}
\label{sec:intro:intro}

Suppose you are given a pair of binary trees $A$ and $B$ and the following problem:
\emph{transform $A$ into $B$ using only right rotations.}
Recall that a right rotation is an operation acting locally on a pair of internal nodes of a binary tree, rearranging them like so:
$$
\imgcenter{\includegraphics[width=1.5cm]{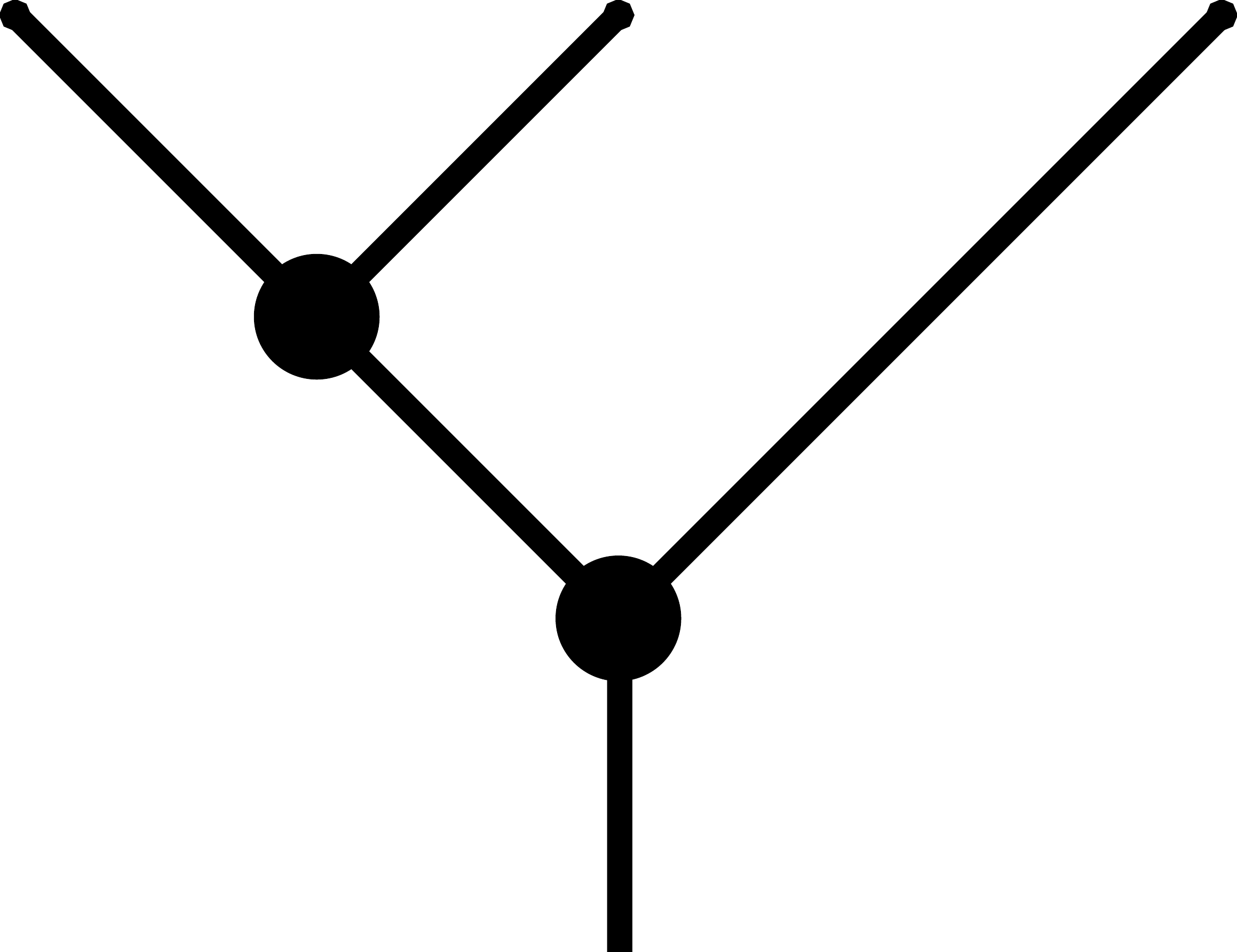}}
 \quad\longrightarrow\quad
\imgcenter{\includegraphics[width=1.5cm]{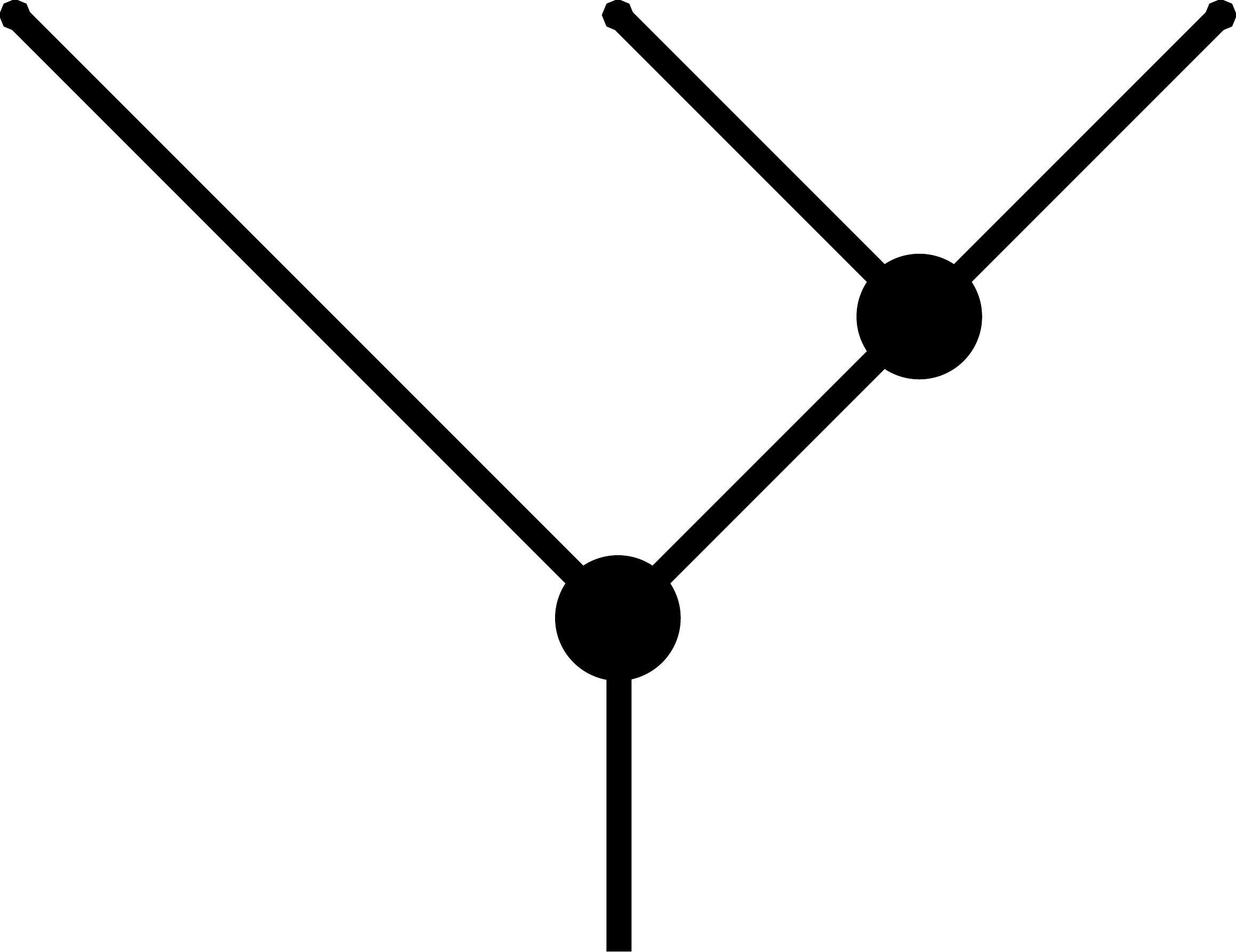}}
$$
Solving this problem amounts to showing that $A \le B$ in the \emph{Tamari order}.
Originally introduced by Dov Tamari in the study of monoids with a partially-defined multiplication operation \cite{Tamari1951phd,Tamari1964,FrTa67}, the Tamari order is the partial ordering on words induced by asking that multiplication obeys a \emph{semi-associative law}\footnote{Clearly, one has to make an arbitrary choice in orienting associativity from left-to-right or right-to-left, and Tamari's original papers in fact took the opposite convention. %
The literature is inconsistent about this, but since the two possible orders defined are strictly dual it does not make much difference. \label{foot1}}
$$
(A*B)*C \le A*(B*C)
$$
and is monotonic in each argument:
$$
\infer{A*B \le A'*B}{A \le A'}
\quad\quad
\infer{A*B \le A*B'}{B \le B'}
$$
For example, the word $(p * (q*r))*s$ is below the word $p * (q * (r * s))$ in the Tamari order.
The variables $p,q,\dots$ are just placeholders and what really matters is the underlying shape of such ``fully-bracketed words'', which is what justifies the above description of the Tamari order in terms of tree rotations.
Since, binary trees are enumerated by the ubiquitous Catalan numbers (there are $\Catalan{n} = \binom{2n}{n}/(n+1)$ distinct binary trees with $n$ internal nodes) which also count many other isomorphic families of objects, the Tamari order has many other equivalent formulations as well, such as on strings of balanced parentheses \cite{HuTa72}, triangulations of a polygon \cite{STT88},  or Dyck paths \cite{BeBo2009} (see also \cite[pp.~474--475]{KnuthTACP4}).

\begin{figure}
\begin{center}
\includegraphics[width=8cm]{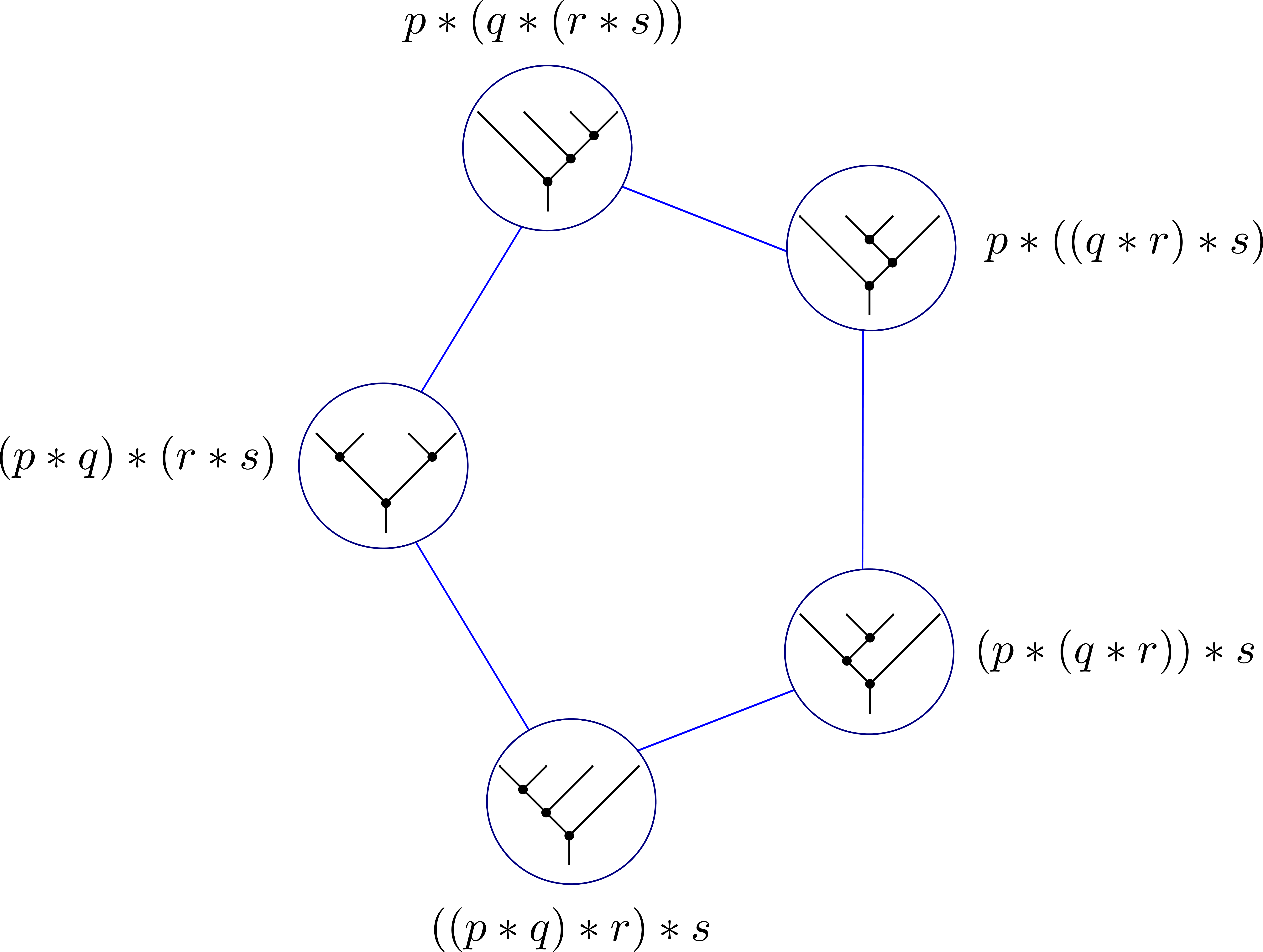} 
\caption{The Tamari lattice $\Tam{3}$.}
\label{fig:tam3}
\end{center}
\end{figure}

For any fixed natural number $n$, the $\Catalan{n}$ objects of that size form a lattice under the Tamari order, which is called the \emph{Tamari lattice} $\Tam{n}$.
For example, Figure~\ref{fig:tam3} shows the Hasse diagram of $\Tam{3}$, which has the shape of a pentagon, and readers familiar with category theory may recognize this as ``Mac~Lane's pentagon'' \cite{MacLaneCWM}.
More generally, a fascinating property of the Tamari order is that each lattice $\Tam{n}$ generates via its Hasse diagram the underlying graph of an $(n-1)$-dimensional polytope called an ``associahedron'' \cite{Stasheff1963,TamariFestschrift}.

\subsection{A Lambekian analysis of the Tamari order}
\label{sec:intro:lambek}

In this paper we will consider a surprisingly elementary (but to the best of my knowledge previously unstudied) presentation of the Tamari order as a \emph{sequent calculus} in the spirit of Lambek \cite{Lambek1958,Lambek1961}.
This calculus consists of just one \emph{left rule} and one \emph{right rule:}
$$
\infer[*L]{\tseq{A*B,\Delta}{C}}{\tseq{A,B,\Delta}{C}}
\qquad
\infer[*R]{\tseq{\Gamma,\Delta}{A*B}}{\tseq{\Gamma}{A} & \tseq{\Delta}{B}}
$$
together with two \emph{structural rules:}
$$
\infer[id]{\tseq{A}{A}}{}
\qquad
\infer[cut]{\tseq{\Gamma,\Theta,\Delta}{B}}{\tseq{\Theta}{A} & \tseq{\Gamma,A,\Delta}{B}}
$$
Here, $\Gamma$, $\Delta$, and $\Theta$ range over lists of formulas called \emph{contexts}, and we write a comma to indicate concatenation of contexts (which is a strictly associative operation).

In fact, all of these rules come straight from Lambek \cite{Lambek1958}, except for the $*L$ rule which is a restriction of his left rule for products.
Lambek's original rule looked like this:
$$\infer[*L^\orig]{\tseq{\Gamma,A*B,\Delta}{C}}{\tseq{\Gamma,A,B,\Delta}{C}}$$
That is, Lambek's left rule allowed the formula $A*B$ to appear anywhere in the context, whereas our more restrictive rule $*L$ requires the formula to appear at the leftmost end of the context.
It turns out that this simple variation makes all the difference for capturing the Tamari order!

For example, here is a sequent derivation of the entailment
$(p * (q*r))*s \le p * (q * (r * s))$
(we write $L$ and $R$ as short for $*L$ and $*R$, and don't bother labelling instances of $id$):
$$
\infer[L]{\tseq{(p * (q*r))*s}{p * (q * (r * s))}}{
\infer[L]{\tseq{p * (q*r),s}{p * (q * (r * s))}}{
\infer[R]{\tseq{p, q*r,s}{p * (q * (r * s))}}{
  \infer{\tseq{p}{p}}{} & 
  \infer[L]{\tseq{q*r,s}{q*(r*s)}}{
  \infer[R]{\tseq{q,r,s}{q*(r*s)}}{
    \infer[R]{\tseq{q,r}{q*r}}{
        \infer{\tseq{q}{q}}{} & 
        \infer{\tseq{r}{r}}{}
    } &
    \infer{\tseq{s}{s}}{}}}}}}
$$
If we had full access to Lambek's original rule then we could also derive the converse entailment (which is false for Tamari):
$$
\infer[L]{\tseq{p * (q * (r * s))}{(p * (q*r))*s}}{
\infer[L^\orig]{\tseq{p, q * (r * s)}{(p * (q*r))*s}}{
\infer[L^\orig]{\tseq{p, q, r * s}{(p * (q*r))*s}}{
\infer[R]{\tseq{p, q, r,s}{(p * (q*r))*s}}{
  \infer[R]{\tseq{p, q,r}{p * (q*r)}}{
   \infer{\tseq{p}{p}}{} & 
    \infer[R]{\tseq{q,r}{q*r}}{
        \infer{\tseq{q}{q}}{} & 
        \infer{\tseq{r}{r}}{}
    }} &
  \infer{\tseq{s}{s}}{}
}}}}
$$
But with the more restrictive rule we can't -- the following soundness and completeness result will be established below.
\begin{claim}
\label{claim:soundcomplete}
$\tseq{A}{B}$ is derivable using the rules $*L$, $*R$, $id$, and $cut$ if and only if $A \le B$ holds in the Tamari order.
\end{claim}
\noindent
As Lambek emphasized, the real power of a sequent calculus comes when it is combined with Gentzen's \emph{cut-elimination} procedure \cite{Gentzen35}.
We will prove the following somewhat stronger form of cut-elimination:
\begin{claim}
\label{claim:focusing}
If $\tseq{\Gamma}{A}$ is derivable using the rules $*L$, $*R$, $id$, and $cut$, then it has a derivation using only $*L$ together with the following restricted forms of $*R$ and $id$: %
$$
\infer[*R^\foc]{\tseq{\Gamma^\irr,\Delta}{A*B}}{\tseq{\Gamma^\irr}{A} & \tseq{\Delta}{B}}
\qquad
\infer[id^\atm]{\tseq{p}{p}}{}
$$
where $\Gamma^\irr$ ranges over contexts that don't have a product formula $C*D$ at their leftmost end.
\end{claim}
\noindent
This is actually a \emph{focusing completeness} result in the sense of Andreoli \cite{Andreoli92},
and we will refer to derivations constructed using only the rules $*L$, $*R^\foc$, and $id^\atm$ as \emph{focused derivations}.
(The above derivation of 
$(p * (q*r))*s \le p * (q * (r * s))$
is an example of a focused derivation.)
A basic analysis of these three rules %
confirms that 
any sequent $\tseq{\Gamma}{A}$ has at most one focused derivation.
\noindent
By combining Claims~\ref{claim:soundcomplete} and \ref{claim:focusing}, we therefore have
\begin{claim}
Every valid entailment in the Tamari order has exactly one focused derivation.
\end{claim}
\noindent
This \emph{coherence theorem} is the main contribution of the paper, and we will see that it has several interesting applications.

\subsection{The surprising combinatorics of Tamari intervals, planar maps, and planar lambda terms}
\label{sec:intro:chapo}

The original impetus for this work came from wanting to better understand an apparent link between the Tamari order and lambda calculus, which was inferred indirectly via their mutual connection to the combinatorics of embedded graphs.

About a dozen years ago, Fréderic Chapoton \cite{Chapoton2006} proved the following surprising formula for the number of \emph{intervals} in the Tamari lattice $\Tam{n}$:
\begin{equation}
\frac{2(4n+1)!}{(n+1)!(3n+2)!} \label{tutte-formula1}
\end{equation}
Here, by an ``interval'' of a partially ordered set we just mean a valid entailment $A \le B$, which can also be identified with the corresponding set of elements $[A,B] = \set{C \mid A \le C \le B}$ (a poset with minimum and maximum elements).
For example, the Tamari lattice $\Tam{3}$ displayed in Figure~\ref{fig:tam3} contains 13 intervals.
Chapoton used generating function techniques to show that (\ref{tutte-formula1}) gives the number of intervals in $\Tam{n}$, and we will explain how the above coherence theorem can be used to give a new proof of this result.
As Chapoton mentions, though, the formula itself did not come out of thin air, but rather was found by querying the \emph{On-Line Encyclopedia of Integer Sequences} (OEIS) \cite{OEIS}.
Formula (\ref{tutte-formula1}) is included in OEIS entry \href{https://oeis.org/A000260}{A000260}, and in fact it was derived over half a century ago by the graph theorist Bill Tutte \cite{Tutte62triangles} for a seemingly unrelated family of objects: it counts the number of (3-connected, rooted) triangulations of the sphere with $3(n+1)$ edges.
The same formula is also known to count other natural families of embedded graphs, and in particular it counts the number of bridgeless rooted planar maps with $n$ edges \cite{WalshLehman75}.\footnote{A \emph{rooted planar map} is a connected graph embedded in the 2-sphere or the plane, with one half-edge chosen as the root. 
A (rooted planar) \emph{triangulation} (dually, trivalent map) is a (rooted planar) map in which every face (dually, vertex) has degree three.
A map is said to be \emph{bridgeless} (respectively, \emph{3-connected}) if it has no edge (respectively, pair of vertices) whose removal disconnects the underlying graph. (Cf.~\cite{LZgraphs}.)} %
Sparked by Chapoton's observation, Bernardi and Bonichon \cite{BeBo2009} found an explicit bijection between intervals of the Tamari order and 3-connected rooted planar triangulations, and quite recently, Fang \cite{Fang2016} has proposed new bijections between these three different families of objects (i.e., between 3-connected rooted planar triangulations, bridgeless rooted planar maps, and Tamari intervals).

Meanwhile, in \cite{ZGcorr}, Alain Giorgetti and I gave a bijection between rooted planar maps and a simple fragment of linear lambda calculus consisting of the $\beta$-normal planar terms.
(Here, ``planarity'' of a lambda term essentially means that the order in which variables are used is fixed following a stack discipline; we will discuss the precise definition of planarity and its potential variations later on.)
As with Chapoton's result, this connection between maps and lambda calculus was found using hints from the OEIS, since the sequence enumerating rooted planar maps was already known 
-- and once again this sequence was first computed by Tutte, who derived another simple closed formula for the number of rooted planar maps with $n$ edges
($\frac{2(2n)!3^n}{n!(n+2)!}$).
It is not difficult to check that the bijection described in \cite{ZGcorr} restricts to a bijection between bridgeless rooted planar maps and $\beta$-normal planar terms \emph{with no closed proper subterms.}
This restriction on a (not necessarily $\beta$-normal or planar) term was called ``indecomposability'' in \cite{Ztrivalent}, where it was used to give a characterization of bridgeless rooted trivalent maps as indecomposable linear lambda terms (and, in turn, to reformulate the Four Color Theorem as a statement about indecomposable planar terms).
In any case, this property of the bijection in \cite{ZGcorr} means that formula (\ref{tutte-formula1}) also enumerates indecomposable $\beta$-normal planar terms by size, and a natural question is whether there is a direct bijection between such terms and intervals of the Tamari order.

As we will explain, in a fairly natural way, every closed (not necessarily $\beta$-normal) indecomposable planar term induces both an \emph{application tree} (describing its underlying applicative structure) and a \emph{binding tree} (describing its underlying binding structure, that is, the matching between abstractions and variables).
Well, it so happens that the binding tree is always below the application tree in the Tamari order!
Trying to prove this fact by induction leads directly to consideration of sequents $\tseq{\Gamma}{A}$, because the binding structure of an \emph{open} indecomposable planar term (with an arbitrary number of free variables) is naturally described as a \emph{list} of trees.
We can then easily build by induction a mapping 
$$
M \quad\mapsto \quad \deduce{\tseq{\bindtrac{M}}{\applica{M}}}{\deriv{D}_M}
$$
from indecomposable planar terms $M$ to derivations $\deriv{D}_M$ showing that the binding forest $\bindtrac{M}$ is below the application tree $\applica{M}$ in the the Tamari order.
Composing with the ``forgetful'' transformation from derivations to their conclusions, we therefore obtain a mapping 
$$M \quad\mapsto\quad (\bindtrac{M},\applica{M})$$
from indecomposable planar terms to (generalized) intervals.
One can show that this mapping is surjective, but not injective.
This is where the coherence theorem comes in: by inspection, the derivation $\deriv{D}_M$ is focused if and only if the term $M$ is $\beta$-normal, and therefore, the mapping $M \mapsto (\bindtrac{M},\applica{M})$ is bijective from indecomposable $\beta$-normal planar terms to Tamari intervals.

\subsection{Outline}

The remainder of the paper is organized as follows.
In Section~\ref{sec:seqcal} we establish all of the proof-theoretic properties claimed above, including soundness and completeness, focusing, and coherence of the sequent calculus for the Tamari order.
In Section~\ref{sec:enumeration} we concisely discuss how the coherence theorem can be applied to give a new proof of formula (\ref{tutte-formula1}) for the number of intervals in $\Tam{n}$, simplifying Chapoton's original proof.
Finally, in Section~\ref{sec:bijection} we recall some basic lambda calculus notions, then turn to the combinatorics of linear lambda terms, and explain how to construct the aforementioned bijection between Tamari intervals and indecomposable $\beta$-normal planar terms.

\section{A sequent calculus for the Tamari order}
\label{sec:seqcal}

\subsection{Definitions and terminology}
\label{sec:seqcal:def}

For reference, we recall here the definition of the sequent calculus introduced in \ref{sec:intro:lambek}, and clarify some notational conventions.
The four rules of the sequent calculus are:
$$
\infer[*L]{\tseq{A*B,\Delta}{C}}{\tseq{A,B,\Delta}{C}}
\qquad
\infer[*R]{\tseq{\Gamma,\Delta}{A*B}}{\tseq{\Gamma}{A} & \tseq{\Delta}{B}}
$$
$$
\infer[id]{\tseq{A}{A}}{}
\qquad
\infer[cut]{\tseq{\Gamma,\Theta,\Delta}{B}}{\tseq{\Theta}{A} & \tseq{\Gamma,A,\Delta}{B}}
$$
Uppercase Latin letters ($A,B,\dots$) range over \definand{formulas}, which can be either \definand{compound} ($A*B$) or \definand{atomic} (ranged over by lowercase Latin letters $p,q,\dots$).
Uppercase Greek letters $\Gamma,\Delta,\dots$ range over \definand{contexts}, which are (possibly empty) lists of formulas, with concatenation of  contexts indicated by a comma.
(Let us emphasize that as in Lambek's system \cite{Lambek1958} but in contrast to Gentzen's original sequent calculus \cite{Gentzen35}, there are no rules of ``weakening'', ``contraction'', or ``exchange'', so the order and the number of occurrences of a formula within a context matters.)
A \definand{sequent} is a pair of a context $\Gamma$ and a formula $A$.

Abstractly, a \definand{derivation} is a tree (formally, a \emph{rooted planar tree with boundary}, cf.~\cite{Kock2011poly-trees}) whose internal nodes are labelled by the names of rules and whose edges are labelled by sequents satisfying the constraints indicated by the given rule.
The \definand{conclusion} of a derivation is the sequent labelling its outgoing root edge, while its \definand{premises} are the sequents labelling any incoming leaf edges.
A derivation with no premises is said to be \definand{closed}.%

We write ``$\tseq{\Gamma}{A}$'' as a notation for sequents, but also sometimes as a shorthand to indicate that the given sequent is \definand{derivable} using the above rules, in other words that there exists a closed derivation whose conclusion is that sequent (it will always be clear which of these two senses we mean).
Sometimes we will need to give an explicit name to a derivation with a given conclusion, in which case we place it over the sequent arrow.

As in \ref{sec:intro:lambek}, when constructing derivations we sometimes write $L$ and $R$ as shorthand for $*L$ and $*R$, and usually don't bother labelling the instances of $id$ and $cut$ since they are clear from context.

Finally, define the \definand{frontier} $\frontier{A}$ of a formula $A$ to be the ordered list of atoms occurring in $A$ (i.e., by $\frontier{A*B} = \frontier{A},\frontier{B}$ and $\frontier{p} = p$), and the frontier of a context $\Gamma = A_1,\dots,A_n$ as the concatenation of frontiers $\frontier{\Gamma} = \frontier{A_1},\dots,\frontier{A_n}$.
The following properties are immediate by examination of the four sequent calculus rules.
\begin{proposition}%
\label{prop:frontier}
Suppose that $\tseq{\Gamma}{A}$. Then
\begin{enumerate}
\item (Refinement:) $\frontier{\Gamma} = \frontier{A}$.
\item (Relabelling:) $\tseq{\sigma\Gamma}{\sigma A}$, where $\sigma$ is any relabelling function on atoms.
\end{enumerate}
\end{proposition}

\subsection{Completeness}
\label{sec:seqcal:completeness}

We begin by establishing completeness relative to the Tamari order, which is the easier direction.

\begin{theorem}[Completeness]
\label{thm:completeness}
If $A \le B$ then $\tseq{A}{B}$.
\end{theorem}
\begin{proof}
We must show that the relation $\tseq{A}{B}$ is reflexive and transitive, and that the multiplication operation satisfies a semi-associative law and is monotonic in each argument.
All of these properties are straightforward:
\begin{enumerate}
\item Reflexivity: immediate by $id$.
\item Transitivity: immediate by $cut$.
\item Semi-associativity: 
$$
\infer[L]{\tseq{(A*B)*C}{A*(B*C)}}{
\infer[L]{\tseq{A*B,C}{A*(B*C)}}{
\infer[R]{\tseq{A,B,C}{A*(B*C)}}{
 \infer{\tseq{A}{A}}{} &
 \infer[R]{\tseq{B,C}{B*C}}{
   \infer{\tseq{B}{B}}{} & \infer{\tseq{C}{C}}{}}}}}
$$
\item Monotonicity:
$$
\infer[L]{\tseq{A*B}{A'*B}}{
\infer[R]{\tseq{A,B}{A'*B}}{\tseq{A}{A'} & \infer{\tseq{B}{B}}{}}}
\qquad
\infer[L]{\tseq{A*B}{A*B'}}{
\infer[R]{\tseq{A,B}{A*B'}}{\infer{\tseq{A}{A}}{} & \tseq{B}{B'}}}
$$
\end{enumerate}
\end{proof}

\subsection{Soundness}
\label{sec:seqcal:soundness}

To prove soundness relative to the Tamari order, first we have to explain the interpretation of general sequents.
The basic idea is that we can interpret a \emph{non-empty context} as a \emph{left-associated product}.
Thus, a general sequent of the form
$$
\tseq{A_1,A_2,\dots,A_n}{B}
$$
(where $n\ge 1$) is interpreted as an entailment of the form
$$
(\cdots(A_1*A_2)*\cdots)*A_n \le B
$$
in the Tamari order.
Visualizing everything in terms of binary trees, the sequent can be interpreted like so:
$$
\vcenter{\hbox{
\begin{tikzpicture}[scale=0.6,thick, grow'=up,
    level 1/.style={level distance=1cm},
    level 2/.style={level distance=1cm,sibling distance=2cm},
    level 3/.style={level distance=1cm,sibling distance=2cm},
    level 4/.style={level distance=1cm,sibling distance=2cm}]
    \coordinate
        child { [fill] circle (4pt)
         child { [fill] circle (4pt)
            child { [fill] circle (4pt)
              child { node {$A_1$} }
              child { node (a2) {$A_2$} }
            }
            child { node (an-1) {$\color{white}A$} }
          }
          child { node (an) {$A_n$} }
        };
    \node at ($ (a2)!0.45!(an) + (0.1,0.1)$) {\rotatebox{0}{$\ddots$}}; 
\end{tikzpicture}}}
\quad\longrightarrow\quad
\vcenter{\hbox{
\begin{tikzpicture}[scale=0.6,thick, grow'=up,
    level 1/.style={level distance=1cm}]
    \coordinate child { node {$B$} };
\end{tikzpicture}}}
$$
That is, the context provides information about the \emph{left-branching spine} of the tree which is below in the Tamari order.

Let $\toF{-}$ be the operation taking any non-empty context $\Gamma$ to a formula $\toF{\Gamma}$ by the above interpretation.
The operation is defined by the following equations:
\begin{align*}
\toF{A} &= A \\
\toF{\Gamma,A} &= \toF{\Gamma} * A
\end{align*}
Critical to soundness of the sequent calculus is the following ``colax'' property of $\toF{-}$:
\begin{proposition}
\label{prop:colax}
$\toF{\Gamma,\Delta} \le \toF{\Gamma}*\toF{\Delta}$ for all non-empty contexts $\Gamma$ and $\Delta$.
\end{proposition}
\begin{proof}
By induction on $\Delta$.
The case of a singleton context $\Delta = A$ is immediate.
Otherwise, if $\Delta = (\Delta',A)$, we have
\begin{align*}
\toF{\Gamma,\Delta',A} &= \toF{\Gamma,\Delta'}*A
\\ &\le (\toF{\Gamma}*\toF{\Delta'})*A
\\ &\le \toF{\Gamma}*(\toF{\Delta'}*A)
\\ & = \toF{\Gamma}*\toF{\Delta',A}
\end{align*}
where the first inequality is by the inductive hypothesis and monotonicity, while the second inequality is by the semi-associative law.
\end{proof}
\noindent
The operation $\toF{-}$ can also be equivalently described in terms of a \emph{right action} $\actF{A}{\Delta}$ of an arbitrary context on a formula, where this action is defined by the following equations:
\begin{align*}
\actF{A}{\cdot} &= A \\
\actF{A}{(\Delta,B)} &= (\actF{A}{\Delta})*B
\end{align*}
We will make use of a few simple properties of $\actF{-}{\Delta}$:
\begin{proposition}
\label{prop:equiv}
$\toF{\Gamma,\Delta} = \actF{\toF{\Gamma}}{\Delta}$ for all non-empty contexts $\Gamma$ and arbitrary contexts $\Delta$.
\end{proposition}
\begin{proposition}[Monotonicity]
\label{prop:monotonic}
If $A \le A'$ then $\actF{A}{\Delta} \le \actF{A'}{\Delta}$.
\end{proposition}
\begin{proof}
Both properties are immediate by induction on $\Delta$, where in the case of Prop.~\ref{prop:monotonic} we apply monotonicity of the operations $-*B$.
\end{proof}

\noindent
We are now ready to prove soundness.
\begin{theorem}[Soundness]
\label{thm:soundness}
If $\tseq{\Gamma}{A}$ then $\toF{\Gamma} \le A$.
\end{theorem}
\begin{proof}
By induction on the (closed) derivation of $\tseq{\Gamma}{A}$.
There are four cases, corresponding to the four rules of the sequent calculus:
\begin{itemize}
\item (Case $*L$): The derivation ends in
$$\infer[*L]{\tseq{A*B,\Delta}{C}}{\tseq{A,B,\Delta}{C}}$$
By induction we have $\toF{A,B,\Delta} \le C$, but by Prop.~\ref{prop:equiv} we have $\toF{A*B,\Delta} = \actF{\toF{A*B}}{\Delta} = \actF{(A*B)}{\Delta} = \actF{\toF{A,B}}{\Delta} = \toF{A,B,\Delta}$.

\item (Case $*R$): The derivation ends in
$$\infer[*R]{\tseq{\Gamma,\Delta}{A*B}}{\tseq{\Gamma}{A} & \tseq{\Delta}{B}}$$
By induction we have $\toF{\Gamma} \le A$ and $\toF{\Delta} \le B$, hence
$$
\toF{\Gamma,\Delta} \le \toF{\Gamma}*\toF{\Delta} \le A*B
$$
where we apply Prop.~\ref{prop:colax} for the first inequality, and monotonicity in both arguments for the second.
\item (Case $id$): Immediate by reflexivity.
\item (Case $cut$): The derivation ends in
$$
\infer[cut]{\tseq{\Gamma,\Theta,\Delta}{B}}{\tseq{\Theta}{A} & \tseq{\Gamma,A,\Delta}{B}}
$$
We can reason like so:
\begin{align*}
\toF{\Gamma,\Theta,\Delta} & = \actF{\toF{\Gamma,\Theta}}{\Delta} \tag{\ref{prop:equiv}}
\\ &\le \actF{(\toF{\Gamma}*\toF{\Theta})}{\Delta}      \tag{\ref{prop:colax} + monotonicity}
\\ &\le \actF{(\toF{\Gamma}*A)}{\Delta}                 \tag{i.h. + monotonicity}
\\ & = \toF{\Gamma,A,\Delta}                            \tag{\ref{prop:equiv}}
\\  &\le B                                              \tag{i.h.}
\end{align*}
\end{itemize}
\end{proof}

\subsection{Focusing completeness}
\label{sec:seqcal:foccomp}

Cut-elimination theorems are a staple of proof theory, and often provide a rich source of information about a given logic.
In this section we will prove a \emph{focusing completeness} theorem, which is an even stronger form of cut-elimination originally formulated by Andreoli in the setting of linear logic \cite{Andreoli92}.

\begin{definition}\label{def:focusing}
A context $\Gamma$ is said to be \definand{reducible} if its leftmost formula is compound, and \definand{irreducible} otherwise.
A sequent $\tseq{\Gamma}{A}$ is said to be:
\begin{itemize}
\item \definand{left-inverting} if $\Gamma$ is reducible;
\item \definand{right-focusing} if $\Gamma$ is irreducible and $A$ is compound;
\item \definand{atomic} if $\Gamma$ is irreducible and $A$ is atomic.
\end{itemize}
\end{definition}
\begin{proposition}\label{prop:seqclass}
Any sequent is either left-inverting, right-focusing, or atomic.
\end{proposition}
\begin{definition}
A closed derivation $\mathcal{D}$ is said to be \definand{focused} if left-inverting sequents only appear as the conclusions of $*L$, right-focusing sequents only as the conclusions of $*R$, and atomic sequents only as the conclusions of $id$.
\end{definition}
\noindent
We write ``$\Gamma^\irr$'' to indicate that a context $\Gamma$ is irreducible.
\begin{proposition}\label{prop:equivfocus}
A closed derivation is focused if and only if it is constructed using only $*L$ and the following restricted forms of $*R$ and $id$ (and no instances of $cut$):
$$
\infer[*R^\foc]{\tseq{\Gamma^\irr,\Delta}{A*B}}{\tseq{\Gamma^\irr}{A} & \tseq{\Delta}{B}}
\qquad
\infer[id^\atm]{\tseq{p}{p}}{}
$$
\end{proposition}
\noindent
\begin{example}
One way to derive 
$$\tseq{((p*q)*r)*s}{p*((q*r)*s)}$$
is by cutting together the two derivations
$$
 \infer[L]{\tseq{((p*q)*r)*s}{(p*(q*r))*s}}{
  \infer[R]{\tseq{(p*q)*r,s}{(p*(q*r))*s}}{
   \deduce{\tseq{(p*q)*r}{p*(q*r)}}{\mathcal{SA}_{p,q,r}} &
   \infer{\tseq{s}{s}}{}}}
$$
and
$$
 \deduce{\tseq{(p*(q*r))*s}{p*((q*r)*s)}}{\mathcal{SA}_{p,q*r,s}}
$$
where $\mathcal{SA}_{A,B,C}$ is the derivation of the semi-associative law $\tseq{(A*B)*C}{A*(B*C)}$ from the proof of Theorem~\ref{thm:completeness}.
Clearly this is \emph{not} a focused derivation (besides the $cut$ rule, it also uses instances of $*R$ and $id$ with a left-inverting conclusion).
However, it is possible to give a focused derivation of the same sequent:
$$
\infer[L]{\tseq{((p*q)*r)*s}{p*((q*r)*s)}}{
\infer[L]{\tseq{(p*q)*r,s}{p*((q*r)*s)}}{
\infer[L]{\tseq{p*q,r,s}{p*((q*r)*s)}}{
\infer[R]{\tseq{p,q,r,s}{p*((q*r)*s)}}{
  \infer{\tseq{p}{p}}{} &
  \infer[R]{\tseq{q,r,s}{(q*r)*s}}{
   \infer[R]{\tseq{q,r}{q*r}}{
     \infer{\tseq{q}{q}}{} &
     \infer{\tseq{r}{r}}{}} &
  \infer{\tseq{s}{s}}{}}}}}}
$$
\end{example}
\noindent
In the below, we write ``$\fseq{\Gamma}{A}$'' as a shorthand notation to indicate that $\tseq{\Gamma}{A}$ has a (closed) focused derivation, and ``$\deriv{D} : \fseq{A}{B}$'' to indicate that $\deriv{D}$ is a particular focused derivation of $\tseq{A}{B}$.
\begin{theorem}[Foc. comp'ness]
\label{thm:foc-completeness}
If $\tseq{\Gamma}{A}$ then $\fseq{\Gamma}{A}$.
\end{theorem}
\noindent
To prove the focusing completeness theorem, it suffices to show that the $cut$ rule as well as the unrestricted forms of $id$ and $*R$ are all \emph{admissible} for focused derivations, in the proof-theoretic sense that given focused derivations of their premises, we can obtain a focused derivation of their conclusion.
We begin by proving a \emph{focused deduction} lemma (cf.~\cite{ReedPfenning10}), which entails the admissibility of $id$, then show $cut$ and $*R$ in turn.
\begin{lemma}[Deduction]
\label{lem:focgenid}
If $\fseq{\Gamma^\irr}{A}$ implies $\fseq{\Gamma^\irr,\Delta}{B}$ for all $\Gamma^\irr$, then $\fseq{A,\Delta}{B}$.
In particular, $\fseq{A}{A}$.
\end{lemma}
\begin{proof}%
By induction on the formula $A$:
\begin{itemize}
\item (Case $A = p$): Immediate by assumption, taking $\Gamma = p$ and $\tseq{p}{p}$ derived by the $id^\atm$ rule.
\item (Case $A = A_1*A_2$): 
By composing with the $*L$ rule,
$$
\infer[*L]{\tseq{A_1*A_2,\Delta}{B}}{\tseq{A_1,A_1,\Delta}{B}}
$$
we reduce the problem to showing $\fseq{A_1,A_2,\Delta}{B}$, and by the i.h. on $A_1$ it suffices to show that $\fseq{\Gamma_1^\irr}{A_1}$ implies $\fseq{\Gamma^\irr_1,A_2,\Delta}{B}$ for all contexts $\Gamma^\irr_1$.
Let $\deriv{D}_1 : \fseq{\Gamma^\irr_1}{A_1}$.
We can derive $\tseq{\Gamma^\irr_1,A_2}{A_1*A_2}$ by
$$
\deriv{D} \quad=\quad
\infer[*R]{\tseq{\Gamma^\irr_1,A_2}{A_1*A_2}}{
 \deduce{\tseq{\Gamma^\irr_1}{A_1}}{\deriv{D}_1} & \deduce{\tseq{A_2}{A_2}}{\deriv{D}_2}}
$$
where we apply the i.h. on $A_2$ to obtain $\deriv{D}_2$.
Finally, applying the assumption %
to $\deriv{D}$ (with $\Gamma^\irr = \Gamma^\irr_1,A_2$) we obtain the desired derivation of $\tseq{A_1,A_2,\Delta}{B}$.
\end{itemize}
\end{proof}
\noindent
\begin{lemma}[Cut]
\label{lem:cut}
If $\fseq{\Theta}{A}$ and $\fseq{\Gamma,A,\Delta}{B}$ then $\fseq{\Gamma,\Theta,\Delta}{B}$.
\end{lemma}

\begin{proof}
Let $\deriv{D} : \fseq{\Theta}{A}$ and $\deriv{E} : \fseq{\Gamma,A,\Delta}{B}$.
We proceed by a lexicographic induction, first on the cut formula $A$ and then on the pair of derivations $(\deriv{D},\deriv{E})$ (i.e., at each inductive step of the proof, either $A$ gets smaller, or it stays the same as one of $\deriv{D}$ or $\deriv{E}$ gets smaller while the other stays the same).

In the case that $A = p$ we can apply the ``frontier refinement'' property (Prop.~\ref{prop:frontier}) to deduce that $\Theta = p$, so the cut is trivial and we just reuse the derivation $\deriv{E} : \fseq{\Gamma,p,\Delta}{B}$.
Otherwise we have $A = A_1*A_2$ for some $A_1,A_2$, and we proceed by case-analyzing the root rule of $\deriv{E}$:
\begin{itemize}
\item (Case $id^\atm$): Impossible since $A$ is non-atomic.
\item (Case $*R^\foc$): This case splits in two possibilities:
\begin{enumerate}
\item $\exists \Delta_1,\Delta_2$ such that $\Delta = \Delta_1,\Delta_2$ and
$$
\deriv{E}\quad=\quad\infer[*R]{\tseq{\Gamma^\irr,A,\Delta_1,\Delta_2}{B_1*B_2}}{\deduce{\tseq{\Gamma^\irr,A,\Delta_1}{B_1}}{\deriv{E}_1} & \deduce{\tseq{\Delta_2}{B_2}}{\deriv{E}_2}}
$$
\item $\exists \Gamma^\irr_1,\Gamma_2$ such that $\Gamma^\irr =\Gamma^\irr_1,\Gamma_2$ and %
$$
\deriv{E}\quad=\quad\infer[*R]{\tseq{\Gamma^\irr_1,\Gamma_2,A,\Delta}{B_1*B_2}}{\deduce{\tseq{\Gamma^\irr_1}{B_1}}{\deriv{E}_1} & \deduce{\tseq{\Gamma_2,A,\Delta}{B_2}}{\deriv{E}_2}}
$$
\end{enumerate}
In the first case, we cut $\deriv{D}$ with $\deriv{E}_1$ to obtain $\fseq{\Gamma^\irr,\Theta,\Delta_1}{B_1}$, then recombine that with $\deriv{E}_2$ using $*R^\foc$ to obtain $\fseq{\Gamma^\irr,\Theta,\Delta_1,\Delta_2}{B_1*B_2}$.
The second case is similar.

\item (Case $*L$): This case splits into two possibilities:
\begin{enumerate}
\item $\exists C_1,C_2,\Gamma'$ such that $\Gamma = C_1*C_2,\Gamma'$ and
$$\deriv{E}\quad=\quad\infer[*L]{\tseq{C_1*C_2,\Gamma',A,\Delta}{B}}{\deduce{\tseq{C_1,C_2,\Gamma',A,\Delta}{B}}{\deriv{E}'}}$$
We cut $\deriv{D}$ into $\deriv{E}'$ and reapply the $*L$ rule.
\item $\Gamma = \cdot$ and
$$\deriv{E}\quad=\quad\infer[*L]{\tseq{A_1*A_2,\Delta}{B}}{\deduce{\tseq{A_1,A_2,\Delta}{B}}{\deriv{E}'}}$$
We further analyze the root rule of $\deriv{D}$:
\begin{itemize}
\item (Case $*L$): $\exists C_1,C_2,\Theta'$ s.t. $\Theta = C_1*C_2,\Theta'$ and %
$$
\deriv{D}\quad=\quad\infer[*L]{\tseq{C_1*C_2,\Theta'}{A_1*A_2}}{\deduce{\tseq{C_1,C_2,\Theta'}{A_1*A_2}}{\deriv{D}'}}
$$
We cut $\deriv{D}'$ into $\deriv{E}$ and reapply the $*L$ rule.
\item (Case $id^\atm$): Impossible.
\item (Case $*R^\foc$): $\exists \Theta^\irr_1,\Theta_2$ s.t. $\Theta^\irr = \Theta^\irr_1,\Theta_2$ and
$$
\deriv{D}\quad=\quad\infer[*R]{\tseq{\Theta^\irr_1,\Theta_2}{A_1*A_2}}{\deduce{\tseq{\Theta^\irr_1}{A_1}}{\deriv{D}_1} & \deduce{\tseq{\Theta_2}{A_2}}{\deriv{D}_2}}
$$
We cut both $\deriv{D}_1$ and $\deriv{D}_2$ into $\deriv{E}'$ (the cuts are at smaller formulas so the order doesn't matter).

\end{itemize}
\end{enumerate}

\end{itemize}
\end{proof}
\noindent
\begin{lemma}[$*R$ admiss.]
\label{lem:*r}
If $\fseq{\Gamma}{A}$ and $\fseq{\Delta}{B}$ then $\tseq{\Gamma,\Delta}{A*B}$.
\end{lemma}
\begin{proof}
We can derive $\fseq{A,B}{A*B}$ using two instances of the deduction lemma together with the $*R$ rule.
Then we obtain $\fseq{\Gamma,\Delta}{A*B}$ using two instances of $cut$.
\end{proof}
\begin{proof}[Proof of Theorem~\ref{thm:foc-completeness}]
An arbitrary closed derivation can be turned into a focused one by starting at the top of the derivation tree and using the above lemmas %
to interpret any instance of the $cut$ rule and of the unrestricted forms of $id$ and $*R$.
\end{proof}
\noindent
Finally, we mention two simple applications of the focusing completeness theorem.
\begin{proposition}[Frontier invariance]
\label{prop:frontier-inv}
Let $\sigma$ be any relabelling function on atoms.
Then $\tseq{\Gamma}{A}$ if and only if $\frontier{\Gamma} = \frontier{A}$ and $\tseq{\sigma\Gamma}{\sigma A}$.
\end{proposition}
\begin{proof}
The forward direction is Prop.~\ref{prop:frontier}.
For the backward direction we use induction on focused derivations, which is justified by Theorem~\ref{thm:foc-completeness}.
The only interesting case is $*R^\foc$, where we can assume $\frontier{\Gamma,\Delta} = \frontier{A*B}$ and $\fseq{\sigma\Gamma}{\sigma A}$ ($\sigma\Gamma$ irreducible) and $\fseq{\sigma\Delta}{\sigma B}$.
By Prop.~\ref{prop:frontier} we have $\frontier{\sigma\Gamma} = \frontier{\sigma A}$ and $\frontier{\sigma\Delta} = \frontier{\sigma B}$, but then elementary properties of lists imply that $\frontier{\Gamma} = \frontier{A}$ and $\frontier{\Delta} = \frontier{A}$, from which $\fseq{\Gamma}{A}$ ($\Gamma$ irreducible) and $\fseq{\Delta}{B}$ follow by the induction hypothesis, hence $\fseq{\Gamma,\Delta}{A*B}$.
\end{proof}
\noindent
If we let $\sigma = \textunderscore \mapsto p$ be any constant relabelling function, then speaking in terms of the Tamari order, Proposition~\ref{prop:frontier-inv} says that to check that two ``fully-bracketed words'' (a.k.a., formulas) are related, it suffices to check that their frontiers are equal and that the unlabelled binary trees describing their underlying multiplicative structure are related.
Although this fact is intuitively obvious, trying to prove it directly by induction on general derivations fails, because in the case of the $cut$ rule we cannot assume anything about the frontier of the cut formula $A$.

\begin{definition}
\label{defn:maxdecomp}
We say that an irreducible context $\Gamma^\irr$ is a \definand{maximal decomposition} of $A$ if $\tseq{\Gamma^\irr}{A}$, and for any other $\Theta^\irr$, $\tseq{\Theta^\irr}{A}$ implies $\tseq{\Theta^\irr}{\toF{\Gamma^\irr}}$.
\end{definition}
\begin{proposition}
\label{prop:decompinv}
If $\Gamma^\irr$ is a maximal decomposition of $A$, then $\tseq{A,\Delta}{B}$ if and only if $\tseq{\Gamma^\irr,\Delta}{B}$.
\end{proposition}
\begin{proof}
The forward direction is by cutting with $\tseq{\Gamma^\irr}{A}$, the backwards direction is by the deduction lemma (\ref{lem:focgenid}) and the universal property of $\Gamma^\irr$.
\end{proof}
\begin{proposition}
\label{prop:decomp}
Let $\toC{A}$ be the irreducible context defined inductively by:
$$
\toC{p} = p \qquad \toC{A*B} = \toC{A},B
$$
Then $\toC{A}$ is a maximal decomposition of $A$.
\end{proposition}
\begin{proof}
We construct $\tseq{\toC{A}}{A}$ by induction on $A$, and prove the universal property of $\toC{A}$ by induction on focused derivations of $\tseq{\Delta}{A}$.
\end{proof}
\begin{proposition}
\label{prop:mudecompmu}
$\toF{\toC{A}} = A$ and $\toC{\toF{\Theta^\irr}} = \Theta^\irr$. %
\end{proposition}
\noindent
The maximal decomposition $\toC{A}$ of $A$ is essentially the same thing as what Chapoton \cite{Chapoton2006} calls a ``décomposition maximale'' of a binary tree.
The logical characterization expressed in Defn.~\ref{defn:maxdecomp} is quite general, though, and is familiar from studies of focusing in other settings (cf.~\cite{Zunityduality}).

\subsection{The coherence theorem}
\label{sec:seqcal:coherence}

We now come to our main result: %
\begin{theorem}[Coherence]
\label{thm:coherence}
Every derivable sequent has exactly one focused derivation.
\end{theorem}
\noindent
Coherence is a direct consequence of focusing completeness and the following lemma:
\begin{lemma}
\label{lem:atmost1}
For any context $\Gamma$ and formula $A$, there is at most one focused derivation of $\tseq{\Gamma}{A}$.
\end{lemma}
\begin{proof}
We proceed by a well-founded induction on sequents, which can be reduced to multiset induction as follows.
Define the \definand{size} $\sz{A}$ of a formula $A$ by
$$\sz{A*B} = 1+\sz{A}+\sz{B} \qquad \sz{p} = 0$$
(That is, $\sz{A}$ counts the number of multiplication operations occurring in $A$.)
Then any sequent $\tseq{A_1,\dots,A_n}{B}$ induces a multisets of size
$\left(\biguplus_{i=1}^n\sz{A_i}\right) \uplus \sz{B}$, 
and at each step of our induction this multiset will decrease in the multiset ordering.

There are three cases:
\begin{itemize}
\item (A left-inverting sequent $\tseq{A*B,\Delta}{C}$): Any focused derivation must end in $*L$, so we apply the i.h. to $\tseq{A,B,\Delta}{C}$.
\item (A right-focusing sequent $\tseq{\Gamma^\irr}{A*B}$): Any focused derivation must end in $*R$, and decide some splitting of the context into contiguous pieces $\Gamma_1^\irr$ and $\Delta_2$.
However, $\Gamma_1^\irr$ and $\Delta_2$ are uniquely determined by frontier refinement ($\frontier{\Gamma_1^\irr} = \frontier{A}$ and $\frontier{\Delta_2} = \frontier{B}$) and the equation $\Gamma^\irr = \Gamma^\irr_1,\Delta_2$.
So, we apply the i.h. to $\tseq{\Gamma_1^\irr}{A}$ and $\tseq{\Delta_2}{B}$.
\item (An atomic sequent $\tseq{\Gamma^\irr}{p}$): The sequent has exactly one focused derivation if and only if $\Gamma^\irr = p$.
\end{itemize}
\end{proof}
\begin{proof}[Proof of Theorem~\ref{thm:coherence}]
By Theorem~\ref{thm:foc-completeness} and Lemma~\ref{lem:atmost1}.
\end{proof}

\subsection{Notes}

The coherence theorem says in a sense that focused derivations provide a canonical representation for intervals of the Tamari order.
Although the representations are quite different, in this respect it seems roughly comparable to the ``unicity of maximal chains'' that was established by Tamari and Friedman as part of their original proof of the lattice property of $\Tam{n}$ \cite{Tamari1964,FrTa67}.
A natural question is whether the sequent calculus can be used to better understand and further simplify the proofs (cf.~\cite{HuTa72} \cite[\S4]{MelliesART6}) of this lattice property.

An easy observation is that one obtains the dual Tamari order (cf.~Footnote~\ref{foot1}) via a dual restriction of Lambek's original rule, in other words by requiring the product formula to appear at the \emph{rightmost} end of the context.
These two forms of product might also be considered in combination with left and right units, or in combination with Lambek's left and right division operations.
Interestingly, Lambek also introduced a fully non-associative version of his original ``syntactic calculus'' \cite{Lambek1961}.

The name ``coherence theorem'' for Theorem~\ref{thm:coherence} is inspired by the terminology from category theory and Mac~Lane's coherence theorem for monoidal categories \cite{MacLane1963}.
Laplaza \cite{Laplaza1972} extended Mac~Lane's coherence theorem to the situation (very close to Tamari's) where there is no monoidal unit and the \emph{associator} $\alpha_{A,B,C} : (A \otimes B)\otimes C \to A \otimes (B \otimes C)$ is only a natural transformation rather than an isomorphism.
(In the presence of units, this gives rise to the notion of a \emph{skew monoidal category} \cite{LackStreet2014}.)
The precise relationship with our coherence theorem remains to be clarified.

\section{Counting intervals in Tamari lattices}
\label{sec:enumeration}

In this section we explain how the coherence theorem can be used to give a new proof of Chapoton's result (mentioned in the Introduction) that the number of intervals in $\Tam{n}$ is given by Tutte's formula (\ref{tutte-formula1}) for planar triangulations.
We will assume some basic familiarity with generating functions (say, as provided by a combinatorics textbook like \cite{FlajoletSedgewick2009}).

The problem of ``counting intervals'' is to compute the cardinality of the set
$$
\Int{n} = \set{(A,B) \in \Tam{n} \times \Tam{n} \mid A \le B}
$$
as a function of $n$.
By the soundness and completeness theorems as well as the frontier invariance property (Prop.~\ref{prop:frontier-inv}), each $\Tam{n}$ is isomorphic as a partial order to the set of formulas $A$ of size $n$ with any fixed frontier of length $n+1$ (remember that a binary tree with $n$ internal nodes has $n+1$ leaves), ordered by sequent derivability. %
By the coherence theorem, the problem of counting intervals can therefore be reduced to the problem of \emph{counting focused derivations}.

This problem lends itself readily to being solved using generating functions.
Consider the generating functions $L(z,x)$ and $R(z,x)$ defined as formal power series $L(z,x) = \sum_{k,n \in \mathbb{N}} \ell_{k,n} x^k z^n$ and $R(z,x) = \sum_{k,n \in \mathbb{N}} r_{k,n} x^k z^n$, where $\ell_{k,n}$ (respectively, $r_{k,n}$) is the number of focused derivations of sequents whose left-hand side is a context (respectively, irreducible context) of length $k$ and whose right-hand side is a formula of size $n$.
(Without loss of generality in this analysis, we assume that all formulas $A$ of size $n$ have a fixed frontier $\frontier{A} = p^{n+1}$.)
We write $L_1(z)$ to denote the coefficient of $x^1$ in $L(z,x)$.
\begin{proposition}
$L_1(z)$ is the ordinary generating function counting Tamari intervals by size.
\end{proposition}
\begin{proof}
The coefficients $\ell_{1,n}$ give the number of focused derivations of sequents of the form $\fseq{A}{B}$, where $\sz{B} = n$ (and hence $\sz{A} = n$), so $\ell_{1,n} = |\Int{n}|$
by Theorem~\ref{thm:coherence}.
\end{proof}
\begin{proposition}\label{prop:genfunc}
$L$ and $R$ satisfy the equations:
\begin{align}
L(z,x) &= \frac{L(z,x) - x L_1(z)}{x} + R(z,x) \label{eqn:L}\\
R(z,x) &= z R(z,x)L(z,x) + x\label{eqn:R}
\end{align}
\end{proposition}
\begin{proof}
The equations are derived directly from the inductive structure of focused derivations:
\begin{itemize}
\item The first summand in (\ref{eqn:L}) corresponds to the contribution from the $*L$ rule, which transforms any $\fseq{A,B,\Gamma}{C}$ into $\fseq{A*B,\Gamma}{C}$.
The context in the premise must have length $\ge 2$ which is why we subtract the $xL_1(x)$ factor, and the context in the conclusion is one formula shorter which is why we divide by $x$.
The second summand is the contribution from irreducible contexts.
\item
The first summand in (\ref{eqn:R}) corresponds to the contribution from the $*R^\foc$ rule, which transforms $\fseq{\Gamma^\irr}{A}$ and $\fseq{\Delta}{B}$ into $\fseq{\Gamma^\irr,\Delta}{A*B}$: the length of the context in the conclusion is the sum of the lengths of $\Gamma^\irr$ and $\Delta$, while the size of $A*B$ is one plus the sum of the sizes of $A$ and $B$, which is why we multiply $R$ and $L$ together and then by an extra factor of $z$.
The second summand is the contribution from $id^\atm : \fseq{p}{p}$. %
\end{itemize}
\end{proof}
\begin{proposition}
\label{prop:L1R1}
$L_1(z) = R(z,1)$.
\end{proposition}
\begin{proof}
This follows immediately from (\ref{eqn:L}), but we can interpret it constructively as well.
The coefficient of $z^n$ in $R(z,1)$ is the formal sum $\sum_k r_{k,n}$, giving the number of focused derivations of sequents whose right-hand side is a formula of size $n$ and whose left-hand side is an irreducible context of arbitrary length.
But by Props.~\ref{prop:decompinv}--\ref{prop:mudecompmu}, the operations $\toF{-}$ and $\toC{-}$ realize a 1-to-1 correspondence between derivable sequents of the form $\tseq{\Gamma^\irr}{B}$ and ones of the form $\tseq{A}{B}$.
\end{proof}
\noindent
After substituting $L_1(z) = R(z,1)$ into (\ref{eqn:L}) and applying a bit of algebra, we obtain another formula for $L$ in terms of a ``discrete difference operator'' acting on $R$:
\begin{equation}
L(z,x) = x\frac{R(z,x) - R(z,1)}{x-1} \label{eqn:L'}
\end{equation}
The recursive (or ``functional'') equations (\ref{eqn:R}) and (\ref{eqn:L'}) can be easily unrolled (preferably using computer algebra software) to compute the first few dozen coefficients of $R$ and $L$:
\begin{align*}
\scriptstyle R(z,x) &\scriptstyle = x + x^2z + (x^2+2x^3)z^2 + (3x^2 + 5x^3 + 5x^4)z^3 + (13x^2 + 20x^3 + \\
&\scriptstyle  21x^4 + 14x^5)z^4 + (68x^2 + 100x^3 + 105x^4 + 84x^5 + 42x^6)z^5 + \dots  \\
\scriptstyle L_1(z) &\scriptstyle = R(z,1) = 1 + z + 3z^2 + 13z^3 + 68z^4 + 399z^5 + 2530z^6 + 16965 z^7 + \dots
\end{align*}
\begin{theorem}[Chapoton \cite{Chapoton2006}]
\label{thm:chapo}
$|\Int{n}| = \frac{2(4n+1)!}{(n+1)!(3n+2)!}$.
\end{theorem}
\begin{proof}
At this point, we can directly appeal to results of Cori and Schaeffer, because equations (\ref{eqn:R}) and (\ref{eqn:L'}) are a special case of the functional equations given in \cite{CoSch2003} for the generating functions of \emph{description trees of type $(a,b)$}, where $a = b = 1$.
Cori and Schaeffer explained how to solve these equations abstractly for $R(z,1)$ using Brown and Tutte's ``quadratic method'', and then how to derive the explicit formula above in the specific case that $a = b = 1$ via Lagrange inversion (essentially as the formula was originally derived by Tutte for planar triangulations).
\end{proof}
\noindent
Let's take a moment to discuss Chapoton's original proof of Theorem~\ref{thm:chapo}, which it should be emphasized is actually not all that different from the one given here.
Chapoton likewise defines a two-variable generating function $\Phi(z,x)$ enumerating intervals in the Tamari lattices $\Tam{n}$, where the parameter $z$ keeps track of $n$, and the parameter $x$ keeps track of the \emph{number of segments along the left border} of the tree at the lower end of the interval.\footnote{Or rather at the upper end, since Chapoton uses the dual convention for orienting the Tamari order (cf.~Footnote~\ref{foot1}).}
By a combinatorial analysis, Chapoton derives the following functional equation for $\Phi$:
\begin{equation}
\Phi(z,x) = x^2 z(1+\Phi(z,x)/x)\left(1 + \frac{\Phi(z,x) - \Phi(z,1)}{x-1}\right) \label{eqn:chapofun}
\end{equation}
He manipulates this equation and eventually solves for $\Phi(z,1)$ as the root of a certain polynomial, from which he derives Tutte's formula (\ref{tutte-formula1}), again by appeal to another result in the paper by Cori and Schaeffer \cite{CoSch2003}.

If we give a bit of thought to these definitions, it is easy to see that the number of segments along the left border of a tree (= formula) $A$ is equal to the length of its maximal decomposition $\toC{A}$ -- meaning that the generating function $\Phi(z,x)$ apparently contains exactly the same information as $R(z,x)$!
There is a small technicality, however, due to the fact that Chapoton only considers the $\Tam{n}$ for $n\ge 1$.
In fact, the two generating functions are related by a small offset (corresponding to the coefficient of $z^0$ in $R(z,x)$):
\begin{equation}
\Phi(z,x) = R(z,x) - x \label{eqn:reparam}
\end{equation}
Indeed, it can be readily verified that equation (\ref{eqn:chapofun}) follows from (\ref{eqn:R}) and (\ref{eqn:L'}), applying the substitution $R(z,x) = x + \Phi(z,x)$.

\section{The interpretation of indecomposable planar lambda terms by Tamari intervals} %
\label{sec:bijection}

\subsection{Preliminaries}

We recall a few basic definitions from lambda calculus \cite{Barendregt1984}.

A \definand{term} (ranged over by uppercase Latin letters $M,N,\dots$) is either a \definand{variable} (ranged over by lowercase Latin letters $x,y,\dots$) or an \definand{application} $M(N)$ (where $M$ and $N$ are terms) or an \definand{abstraction} $\lambda x.M$ (where $x$ is a variable and $M$ is a term).
By syntactic convention, abstractions take scope to the right as far as possible, so that for example ``$\lambda x.x(\lambda y.y)$'' should be read as an abstraction term, whereas ``$(\lambda x.x)(\lambda y.y)$'' is an application term.

We define the \definand{subterms} of a term as follows:
\begin{itemize}
\item $M$ is a subterm of itself
\item any subterm of $M$ or $N$ is a subterm of $M(N)$
\item any subterm of $M$ is a subterm of $\lambda x.M$
\end{itemize}
The subterms of $M$ are said to ``occur'' in $M$.
Among variables occurring within a term, we distinguish \definand{free} variables from \definand{bound} variables: an abstraction term $\lambda x.M$ is said to bind any free occurrences of $x$ in $M$, and otherwise all variables which are not bound by an abstraction are said to be free.
A term with no free variables is said to be \definand{closed}.
Terms are usually considered up to \definand{$\alpha$-conversion}, or renaming of bound variables (e.g., the terms $\lambda x.x$ and $\lambda y.y$ are $\alpha$-equivalent).
We will assume the \definand{Barendregt convention}, which says that all bound variables have distinct names and that no variable is both free and bound (this condition is always possible to achieve via $\alpha$-conversion).

The basic computation rule of lambda calculus is the rule of \definand{$\beta$-reduction},
\begin{equation}\tag{$\beta$}
(\lambda x.M)(N) \to M[N/x]
\end{equation}
where $M[N/x]$ denotes the (``capture-avoiding'') substitution of $N$ for any free occurrences of $x$ in $M$.
The $\beta$-reduction rule can be performed on any subterm, in other words there are also the following ``congruence'' rules:
$$
\infer{M(N) \to M'(N)}{M \to M'}\quad
\infer{M(N) \to M(N')}{N \to N'}\quad
\infer{\lambda x.M \to \lambda x.M'}{M \to M'}
$$
This defines a rewriting system which is confluent (the Church-Rosser theorem), although there exist infinite reduction sequences $M_1 \to M_2 \to \dots$ (since pure lambda calculus is a universal model of computation).
Two terms are said to be \definand{$\beta$-equivalent} $M =^\beta N$ if they both $\beta$-reduce to the same term $M \to P \leftarrow N$ in any number of steps.
A term which cannot be further reduced is said to be \definand{$\beta$-normal}.

An abstraction $\lambda x.M$ is said to be \definand{linear} if the variable $x$ has exactly one free occurrence in $M$.
By extension, a term $N$ is said to be linear if every abstraction subterm of $N$ is linear, and all free variables of $N$ occur exactly once as subterms.
For example, the terms $\lambda x.\lambda y.y(x)$ and $\lambda x.x(\lambda y.y)$ are linear, but the terms $\lambda x.x(x)$ and $\lambda x.\lambda y.y$ are not.

A term is said to be \definand{indecomposable} \cite{Ztrivalent} if it has no closed proper subterms.
For example, the term $\lambda x.\lambda y.y(x)$ is indecomposable, but the term $\lambda x.x(\lambda y.y)$ is not.
\begin{proposition}
\label{prop:indecomp-abs}
A closed indecomposable term is necessarily an abstraction term $\lambda x.M$, where $M$ is indecomposable.
\end{proposition}

\subsection{Application trees and binding diagrams}

Any linear lambda term is naturally associated with a pair of basic combinatorial objects: a binary tree describing its underlying structure of applications, and a \emph{rooted chord diagram} describing the matching between lambda abstractions and variables.
The quickest way of explaining this is with a picture:
$$
\imgcenter{\includegraphics[width=0.4\textwidth]{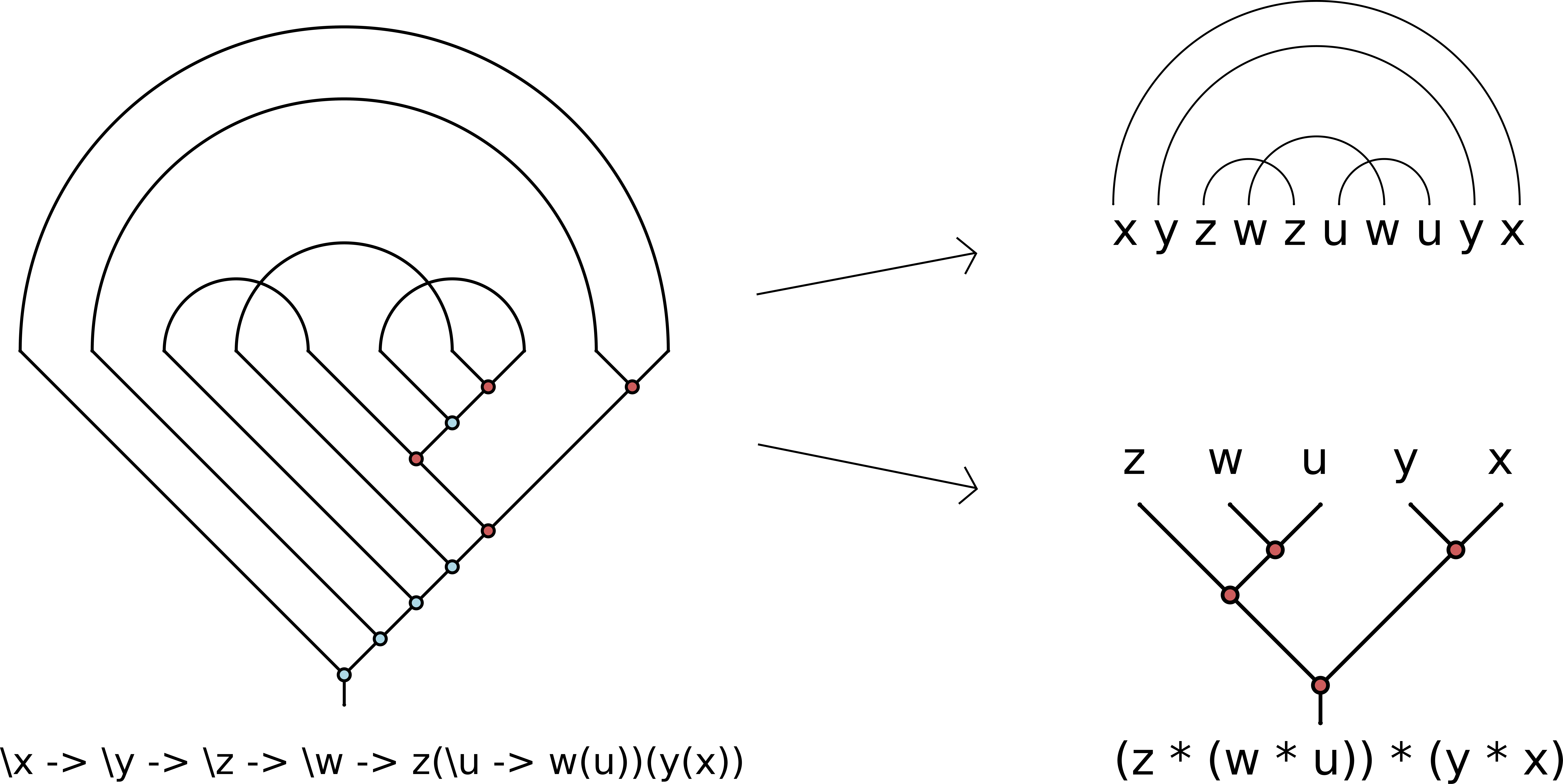}}
$$
On the left we have a diagram that faithfully represents the linear term $\lambda x.\lambda y.\lambda z.\lambda w.z(\lambda u.w(u))(y(x))$ up to $\alpha$-equivalence, with nodes representing applications colored in red and nodes representing abstractions colored in blue.
(This kind of diagrammatic representation is folklore in lambda calculus; for a more thorough discussion, see \cite{Ztrivalent}.)
On the right we've selectively gotten rid of some of the structure of the term to produce two simpler objects.
The \emph{application tree} (depicted in the lower right) collapses abstraction nodes,
keeping only the underlying binary tree of applications (with variables labelling the leaves).
\begin{definition}
\label{defn:apptree}
The \definand{application tree} of a term $M$ is a (leaf-labelled) binary tree $\apptree{M}$, defined by induction as follows:
\begin{align*}
\apptree{x} &= x\\
\apptree{M(N)} &= \apptree{M} * \apptree{N} \\
\apptree{\lambda x.M} &= \apptree{M}
\end{align*}
\end{definition}
\noindent
On the other hand, the \emph{binding diagram} (depicted in the upper right of the picture) collapses application nodes, 
recording only the \emph{order} of successive lambda abstractions and variable occurrences (or ``uses'').
Reading the diagram from left to right, rising arcs correspond to abstractions and falling arcs to uses.
The fact that this is the binding diagram of a \emph{closed linear} term means that every rising arc is met by exactly one falling arc, and we get the classical notion of rooted chord diagram (also sometimes referred to as an ``arc diagram'' \cite{LZgraphs} or ``matching diagram'' \cite{BCFMM2016}), which has many equivalent purely combinatorial representations such as by \emph{double-occurrence words} ($x\,y\,z\,w\,z\,u\,w\,u\,y\,x$) or by \emph{fixed point-free involutions} ($(1\ 10)(2\ 9)(3\ 5)(4\ 7)(6\ 8)$).
More generally, the binding diagram of a linear term with free variables corresponds to a rooted chord diagram \emph{where chords can have unattached ends} (i.e., to an ``open matching'' in the sense of \cite{BCFMM2016}).
Such diagrams can be represented faithfully by ``at-most-double-occurrence'' words (that is, sequences where every symbol occurs either once or twice), or equivalently by involutions with fixed points.
\begin{definition}
\label{defn:bindword}
The \definand{binding diagram} of a linear term $M$ is a sequence $\binddiag{M}$, defined by induction as follows:
\begin{align*}
\binddiag{x} &= x\\
\binddiag{M(N)} &= \binddiag{M}, \binddiag{N} \\
\binddiag{\lambda x.M} &= x, \binddiag{M}
\end{align*}
\end{definition}
\begin{proposition}\label{prop:dow}
If $M$ is linear, then $\binddiag{M}$ is an at-most-double-occurrence word (assuming the Barendregt convention). 
If moreover $M$ is closed, then $\binddiag{M}$ is a double-occurrence word.
\end{proposition}
\noindent
By the isomorphism between rooted chord diagrams and double-occurrence words, we will therefore view the binding diagram of a linear term interchangeably either as an (at-most-)double-occurrence word or as a rooted chord diagram (potentially with unattached chords).
We refer to chords with an unattached end (i.e., single-occurrence letters) as \emph{free chords}, and to chords with both ends attached (i.e., double-occurrence letters) as \emph{full chords}.

Let the \definand{size} $\sz{M}$ of a linear term $M$ be defined here as the number of internal nodes in its application tree $\apptree{M}$. %
\begin{proposition}
\label{prop:treebindsz}
If $M$ is a closed linear term of size $n$, then its binding diagram $\binddiag{M}$ is a rooted chord diagram with $n+1$ full chords (i.e., the corresponding double-occurrence word has $2n+2$ letters).
More generally, if $M$ is a linear term of size $n$ with $k$ free variables, then its binding diagram has $n+1-k$ full chords and $k$ free chords.
\end{proposition}

A rooted chord diagram (with no free chords)
is said to be \emph{indecomposable} if it is not the juxtaposition of two rooted chord diagrams -- the corresponding condition on a double-occurrence word is that it is not the concatenation of two double-occurrence words (cf.~\cite{OdMRencoding,Cori2009}).
\begin{proposition}Let $M$ be a closed linear term. If $M$ is indecomposable (i.e., has no closed proper subterms), then $\binddiag{M}$ is an indecomposable double-occurrence word.
\end{proposition}
\noindent
(Note the converse is false: for example, $\lambda x.(\lambda y.y)x$ is not indecomposable, but has an indecomposable binding diagram.)

\subsection{Planarity and the lambda rotation ($\rho$) rule}

A rooted chord diagram is said to be \emph{planar} if it has no pair of crossing arcs.
This translates to
at-most-double-occurrence words as follows: $\gamma$ is planar if for any $x,\delta,x$ occurring as a subword, $\delta$ is a double-occurrence word.
\begin{definition}
\label{defn:planarity}
We say that a linear term $M$ is \definand{planar} just in case its binding diagram $\binddiag{M}$ is planar.
\end{definition}
\noindent
An important comment about this definition is that the class of terms which are considered planar clearly depends upon the ``layout convention'' we use in drawing diagrams, which is implicit in the definition of the binding diagram $\binddiag{M}$.
The precise notion of planarity we consider here was studied
(in an equivalent formulation)
in \cite{ZGcorr}, where it was called ``LR-planarity'' and contrasted with ``RL-planarity'' (see \cite[\S3.1]{ZGcorr}, and also \cite[\S4]{Ztrivalent}): the latter can be defined by replacing the abstraction case of Defn.~\ref{defn:bindword} by $\binddiag{\lambda x.M} = \binddiag{M},x$.
For example, the term $\lambda x.\lambda y.y(\lambda z.z(x))$ is LR-planar, while the term $\lambda x.\lambda y.x(\lambda z.y(z))$ is RL-planar.
A simple observation about planarity is that for any given underlying ``skeleton'' of abstractions and applications sans variable names (i.e., an expression like ``$\lambda\wild.\lambda\wild.\wild(\lambda \wild.\wild(\wild))$''), there is a unique (up to $\alpha$-equivalence) way of filling in the variable names to produce either an LR-planar term or an RL-planar term.

Even though the difference is seemingly trivial, the asymmetry of the lambda calculus means that these two notions of planarity have very different properties.
Notably, the set of LR-planar terms is \emph{not} closed under $\beta$-reduction, which diagrammatically (under these conventions) corresponds to the following operation:
\begin{align*}\tag{$\beta$}
\imgcenter{\includegraphics[width=1.5cm]{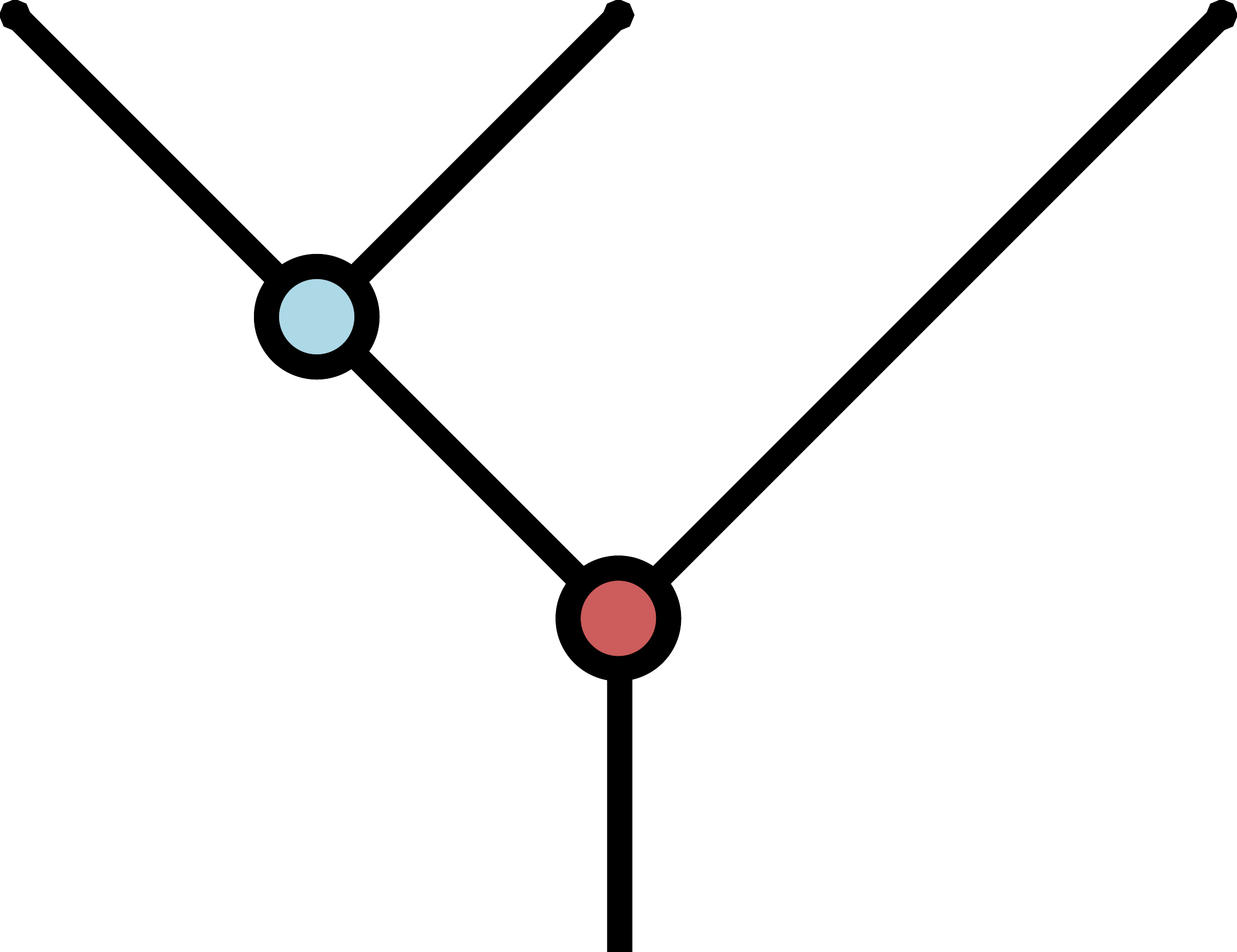}}
 \quad&\longrightarrow\quad
\imgcenter{\includegraphics[width=1.5cm]{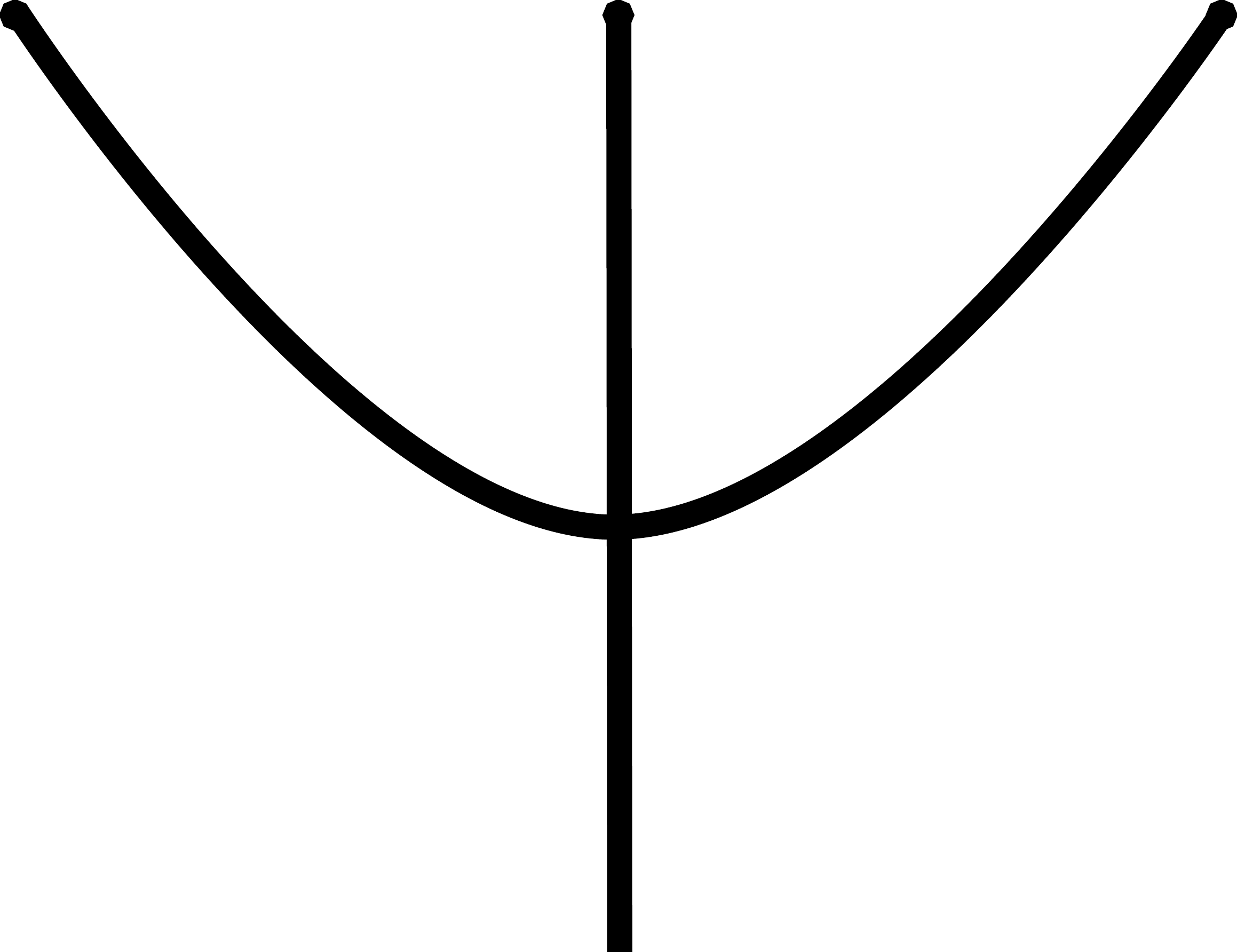}} \\
(\lambda x.M)(N) \quad&\qquad\ \ \quad M[N/x]
\end{align*}
On the other hand, LR-planar terms \emph{are} closed under a ``colored'' version of the right rotation operation, which rotates a lambda abstraction out from the left of an application:
\begin{align*}\tag{$\rho$}
\imgcenter{\includegraphics[width=1.5cm]{betaredLR.pdf}}
 \quad&\longrightarrow\quad 
\imgcenter{\includegraphics[width=1.5cm]{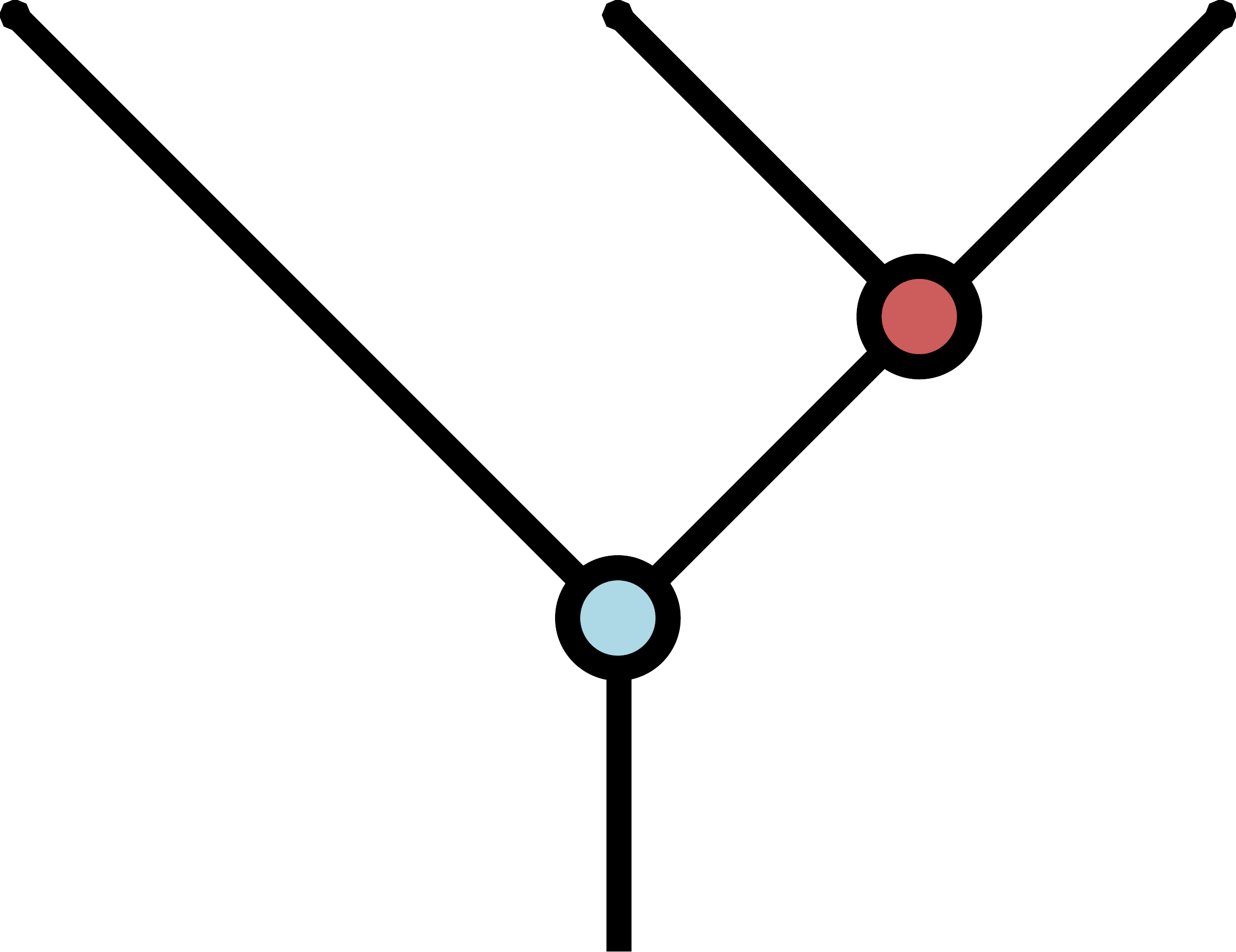}} \\
\quad(\lambda x.M)(N) \quad &\qquad\quad\  \lambda x.M(N)%
\end{align*}
The $\rho$-reduction rule is certainly not a typical lambda calculus rule -- but the amusing coincidence is that the set of $\beta$-normal terms and the set of $\rho$-normal terms coincide.
This is true even though the induced notions of equivalence are very different: let us say that two terms are $\rho$-equivalent ($M =^\rho N$) just in case they both reduce to a common term via some sequence of $\rho$-reductions (applied to any subterms). 
The following observation is key: it says that the pair of the application tree and binding diagram of a linear lambda term form a \emph{complete invariant} for that term up to $\rho$-equivalence.
\begin{theorem}%
For all linear terms $M$ and $N$, $M =^\rho N$ if and only if $\apptree{M} = \apptree{N}$ and $\binddiag{M} = \binddiag{N}$.
\end{theorem}
\begin{proof}
For the forward direction, we check that $\rho$-reduction preserves application trees and binding diagrams, which is evident by inspection of the rule.
For the backward direction, we first verify that every term has a $\rho$-normal form (obvious, since any lambda abstraction can be rotated only finitely many times), then that any two $\rho$-normal forms with equal application trees and binding diagrams must be equal.
Here we can use the fact that $\rho$-normal forms (which, again, are identical with $\beta$-normal forms) have a simple inductive structure.
In particular, we can read off the \emph{head normal form} $\lambda x_1\dots \lambda x_n.((x M_1)\dots M_p)$ of a term by examining the leading arcs of its binding diagram (in this case, $n$ rising arcs $x_1,\dots,x_n$ followed by a falling arc $x$) together with the left-branching spine of its application tree (in this case, with the leftmost leaf labelled $x$, and $p$ subtrees along the right).
Linearity determines how the rest of the binding diagram is partitioned among the subterms $M_1, \dots, M_p$, and we proceed by induction.
\end{proof}

\subsection{From indecomposable planar terms to Tamari intervals}

In this section we at last present the bijection between ($\rho$-/$\beta$-)normal planar indecomposable terms and Tamari intervals, beginning first by explaining how to map any planar indecomposable term to an interval.
It is well-known that rooted planar chord diagrams are enumerated by the Catalan numbers: there are $\Catalan{n}$ rooted planar chord diagrams with $n$ chords, and hence they can be placed in bijection with binary trees with $n$ internal nodes.
Since we were hoping to extract a pair of binary trees and we already have the application tree, that seems like a good start.
Remember, though, that a term with an application tree of size $n$ has $n+1$ chords in its binding diagram (Prop.~\ref{prop:treebindsz}), so there is an apparent indexing mismatch that has to be accounted for. 

\begin{figure}
\begin{center}
\imgcenter{\includegraphics[width=0.4\textwidth]{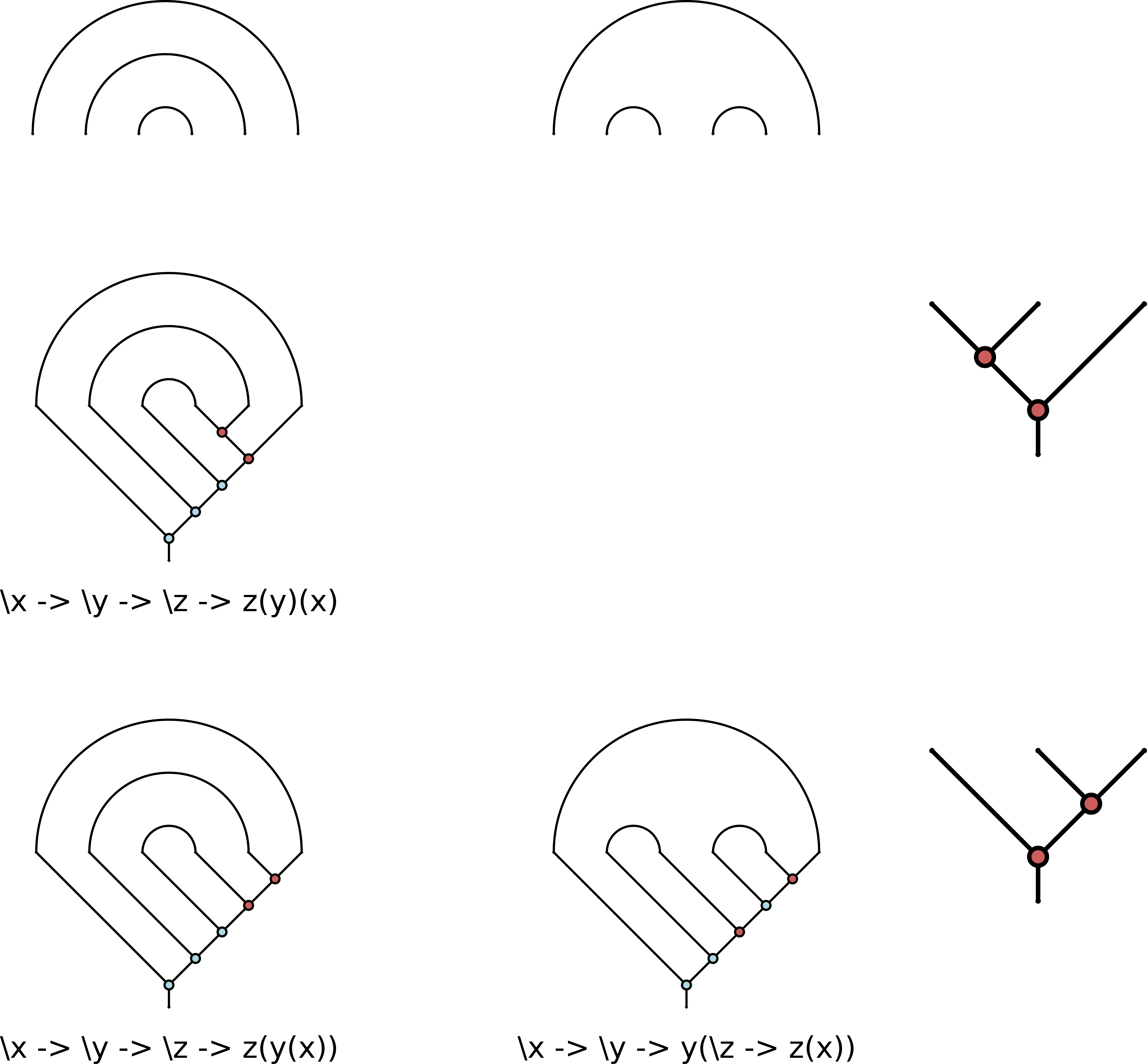}}
\end{center}
\caption{The three indecomposable normal planar terms of size two, organized into rows according to their underlying application trees, and into columns by their underlying binding diagrams.}
\label{fig:lam2}
\end{figure}

\begin{figure}
$$
\imgcenter{\includegraphics[width=0.47\textwidth]{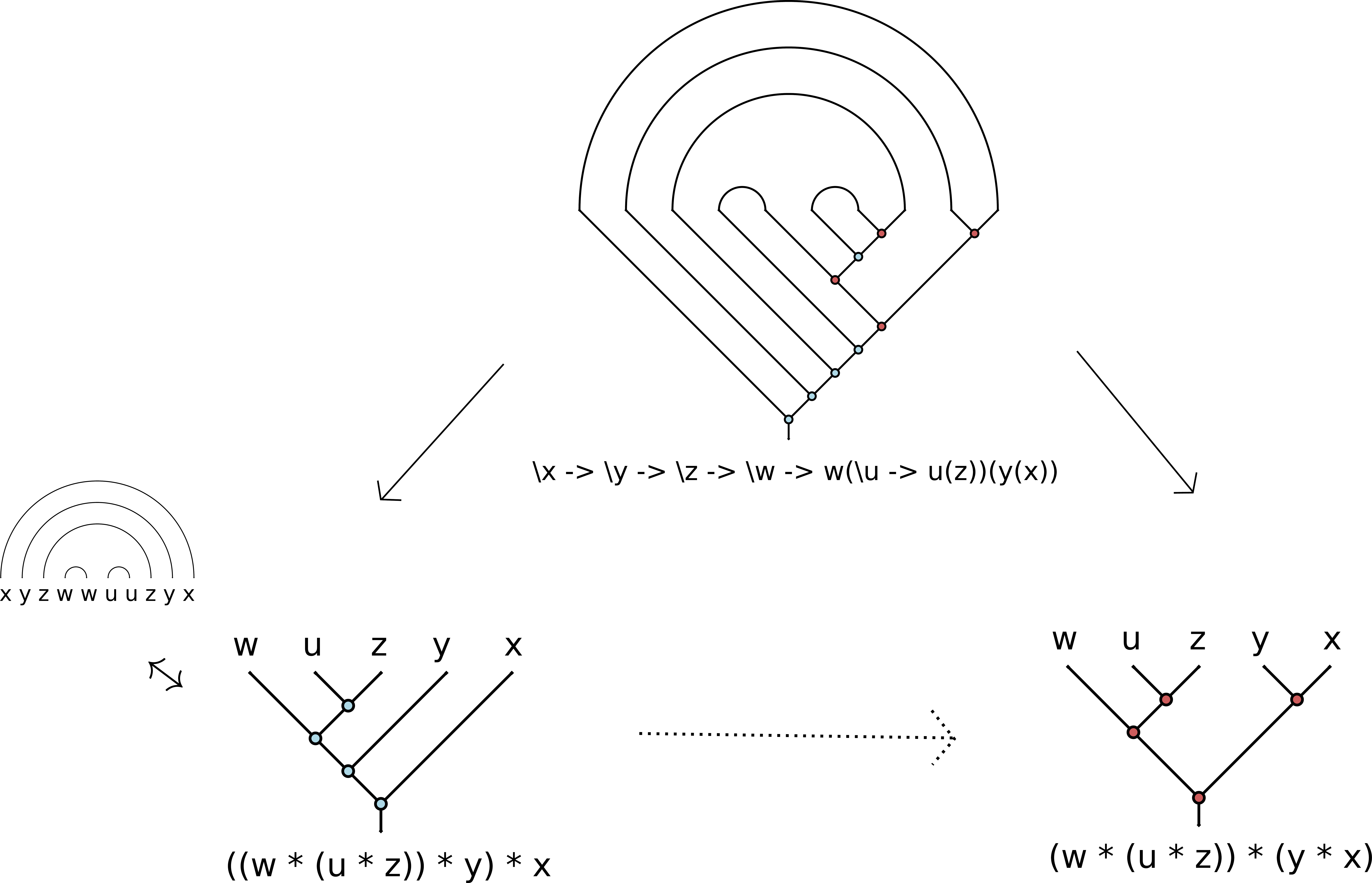}}
$$
\caption{Extracting a Tamari interval from an indecomposable planar term}
\label{fig:lamtam}
\end{figure}

\begin{proposition}
Any indecomposable planar double-occurrence word is necessarily of the form $\beta = x,\gamma,x$, where $\gamma$ is a planar double-occurrence word.
\end{proposition}
\noindent
In other words, every indecomposable rooted planar chord diagram has a unique outermost chord.
This simple property is readily apparent from inspection of the binding diagrams of indecomposable planar terms (see Figures~\ref{fig:lam2}--\ref{fig:lam4-poi}),
and it enables us to transform any indecomposable rooted planar chord diagram with $n+1$ (full) chords into an $n$-node binary tree conveying equivalent information.

Geometrically, the construction could be visualized like so:
$$
\imgcenter{\includegraphics[width=0.47\textwidth]{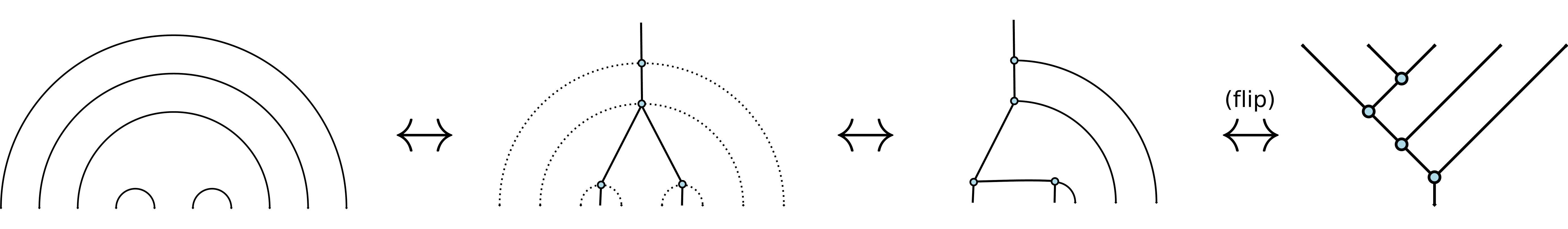}}
$$
Going from left to right, we begin by removing the outermost chord and applying the canonical isomorphism between rooted planar chord diagrams and rooted planar trees
(which interprets chords as nodes, and the covering relation between chords as the parent-child relation between nodes).
Then we apply the (non-canonical) ``left-child, right-sibling'' correspondence between rooted planar trees and binary trees, and flip the resulting binary tree vertically to get our preferred orientation.

The miracle is that whenever we apply this process to the binding diagram $\binddiag{M}$ of a closed indecomposable planar term $M$, we get a tree which is below the application tree $\apptree{M}$ in the Tamari order!
Figure~\ref{fig:lamtam} shows a demonstration of this on a term of size four, and the reader is invited to test it out for herself on any of the diagrams in Figures~\ref{fig:lam2}, \ref{fig:lam3} or \ref{fig:lam4-poi}.

So how do we prove that this always works?
It turns out that the sequent calculus of Section~\ref{sec:seqcal} is especially well-suited (indeed, this was it's original \emph{raison d'être}).
If one tries to give a formal inductive definition of the trees constructed above from the binding diagrams of closed indecomposable planar terms, this quickly leads to the consideration of \emph{binding forests} for open terms -- these correspond to sequent contexts, but they also make sense geometrically, since the binding diagram of an open planar term contains unattached arcs which partition the diagram into disjoint subdiagrams.
\begin{definition}
\label{defn:bindforest}
Let $M$ be any indecomposable planar term with $k$ free variables, for $k \ge 1$.
The \definand{binding forest} of $M$ is a list of $k$ (leaf-labelled) binary trees $\bindfor{M}$, defined by induction as follows:
\begin{align*}
\bindfor{x} &= x \\
\bindfor{M(N)} &= \Gamma,\Delta\quad \text{where }\bindfor{M} = \Gamma\text{ and } \bindfor{N} = \Delta\\
\bindfor{\lambda x.M} &= A*B,\Gamma \quad \text{where } \bindfor{M} = A,B,\Gamma \\
 & \qquad \text{assuming } x\text{ at rightmost end of }\frontier{A}
\end{align*}
By extension, the \definand{binding tree} of a closed indecomposable planar term (cf.~Prop.~\ref{prop:indecomp-abs}) is defined by $\bindtree{\lambda x.M} = \bindfor{M}$.
\end{definition}
\noindent
Note the side-condition on the variable $x$ in the definition of $\bindfor{\lambda x.M}$, which enforces planarity.
\begin{proposition}
If $M$ is an indecomposable planar term with $k\ge 1$ free variables, then $\bindfor{M}$ is well-defined, and the leftmost free variable of $M$ occurs at the rightmost end of the frontier of the first formula in $\bindfor{M}$. 
\end{proposition}
\noindent
With this definition of binding forests as contexts, verifying the ``miracle'' becomes a routine exercise in sequent calculus.
\begin{proposition}
\label{prop:ipttam}
$\tseq{\bindfor{M}}{\apptree{M}}$ for all indecomposable planar terms $M$.
\end{proposition}
\begin{proof}
Immediate by induction on $M$, interpreting abstractions by the $*L$ rule, applications by the $*R$ rule, and variables by the $id$ rule (in its atomic instance $id^\atm$).
\end{proof}
\begin{corollary}
For all closed indecomposable planar terms $M$, the binding tree $\bindtree{M}$ is below the application tree $\apptree{M}$ in the Tamari order.
\end{corollary}
\noindent
Finally, why does this give a \emph{bijection} from closed $\beta$-normal planar indecomposable terms to Tamari intervals?
Well, by examination of the proof of Prop.~\ref{prop:ipttam}, it is clear that we can also go in the opposite direction and turn any sequent calculus derivation involving the rules $*L$, $*R$, and $id^\atm$ back into an indecomposable planar term: just interpret $*L$ as an abstraction (binding the atom at the rightmost end of the frontier of the first formula of the context in the premise), $*R$ as an application, and $id^\atm$ as a variable occurrence.
(The frontier invariance property can be used to rename atoms as necessary, if we want to obtain a term satisfying the Barendregt convention.)
Since such derivations include all the focused ones, this establishes that the mapping from closed indecomposable planar terms to Tamari intervals is surjective, by the focusing completeness theorem.
Moreover, within this fragment of the sequent calculus, the only possible source of a focusing violation is if we have a derivation whose root is $*L$ as the left premise of a derivation whose root is $*R$ -- but such violations can be resolved by permuting $*L$ under $*R$:
$$\scriptsize
\vcenter{\hbox{
\infer[R]{\tseq{A*B,\Gamma_1,\Gamma_2}{C*D}}{
  \infer[L]{\tseq{A*B,\Gamma_1}{C}}{\tseq{A,B,\Gamma_1}{C}} &
  \tseq{\Gamma_2}{D}}
}}
\to
\vcenter{\hbox{
\infer[L]{\tseq{A*B,\Gamma_1,\Gamma_2}{C*D}}{
\infer[R]{\tseq{A,B,\Gamma_1,\Gamma_2}{C*D}}{
 \tseq{A,B,\Gamma_1}{C} & \tseq{\Gamma_2}{D}}}
}}
$$
This transformation on derivations corresponds precisely to performing a $\rho$-reduction on the corresponding lambda terms.
Therefore focused derivations correspond precisely to $\rho$-normal indecomposable planar terms, 
but the latter also happen be $\beta$-normal, and by the coherence theorem we have:
\begin{theorem}
Closed indecomposable $\beta$-normal planar terms with $n$ applications are in bijection with intervals in $\Tam{n}$.
There are exactly $\frac{2(4n+1)!}{(n+1)!(3n+2)!}$ of them.
\end{theorem}

\section*{Acknowledgment}

I thank Tarmo Uustalu for telling me about skew monoidal categories.
The diagrams in this paper were created using the help of Brent Yorgey's Diagrams library for Haskell. %

\bibliographystyle{IEEEtranS}
\bibliography{tamari}

\onecolumn

\appendix

\begin{figure}[h]
\begin{center}
\imgcenter{\includegraphics[width=\textwidth]{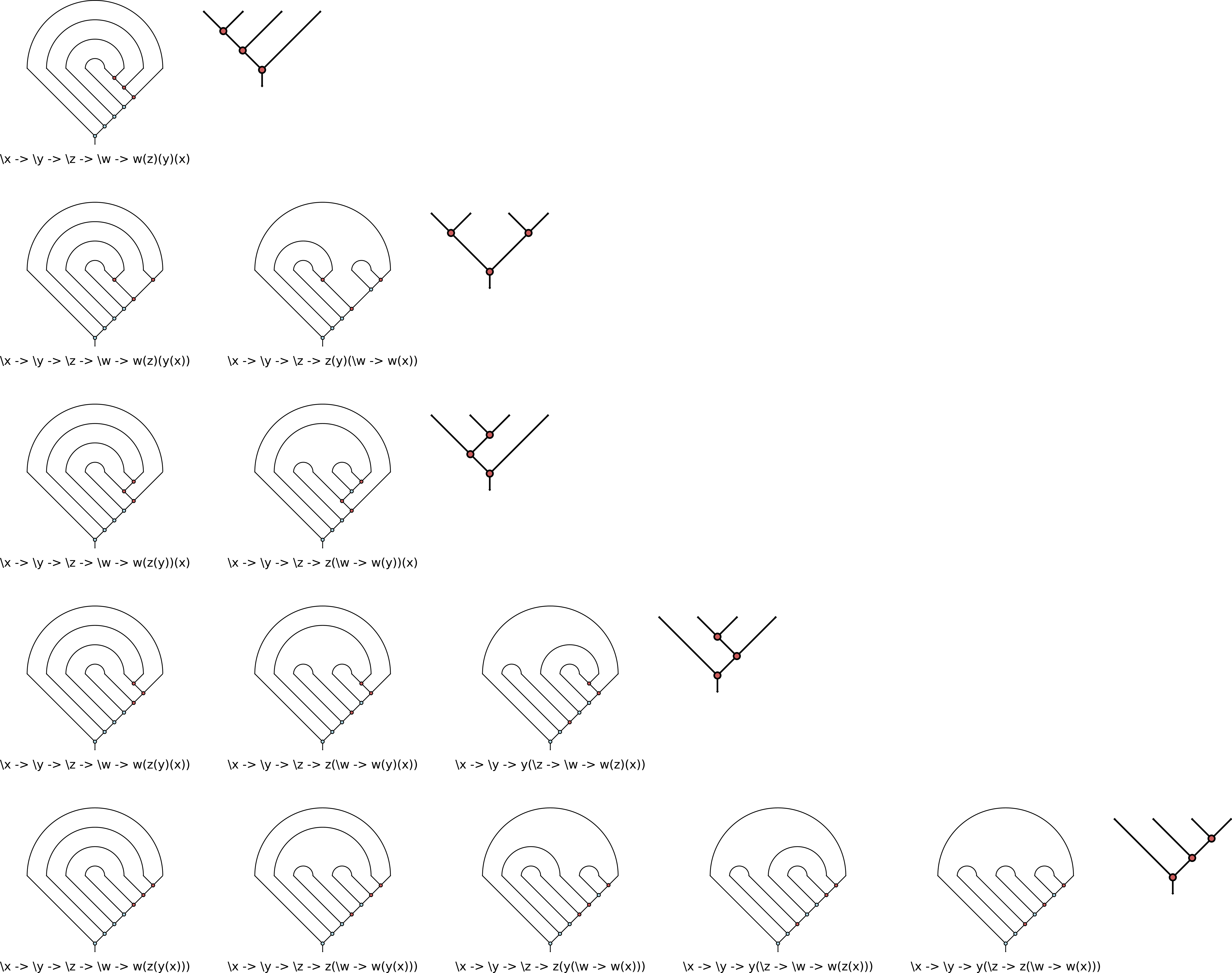}}
\caption{All 13 indecomposable normal planar terms of size three, organized into rows according to their underlying application tree.}
\label{fig:lam3}
\end{center}
\end{figure}

\begin{figure}[h]
\begin{center}
\imgcenter{\includegraphics[width=\textwidth]{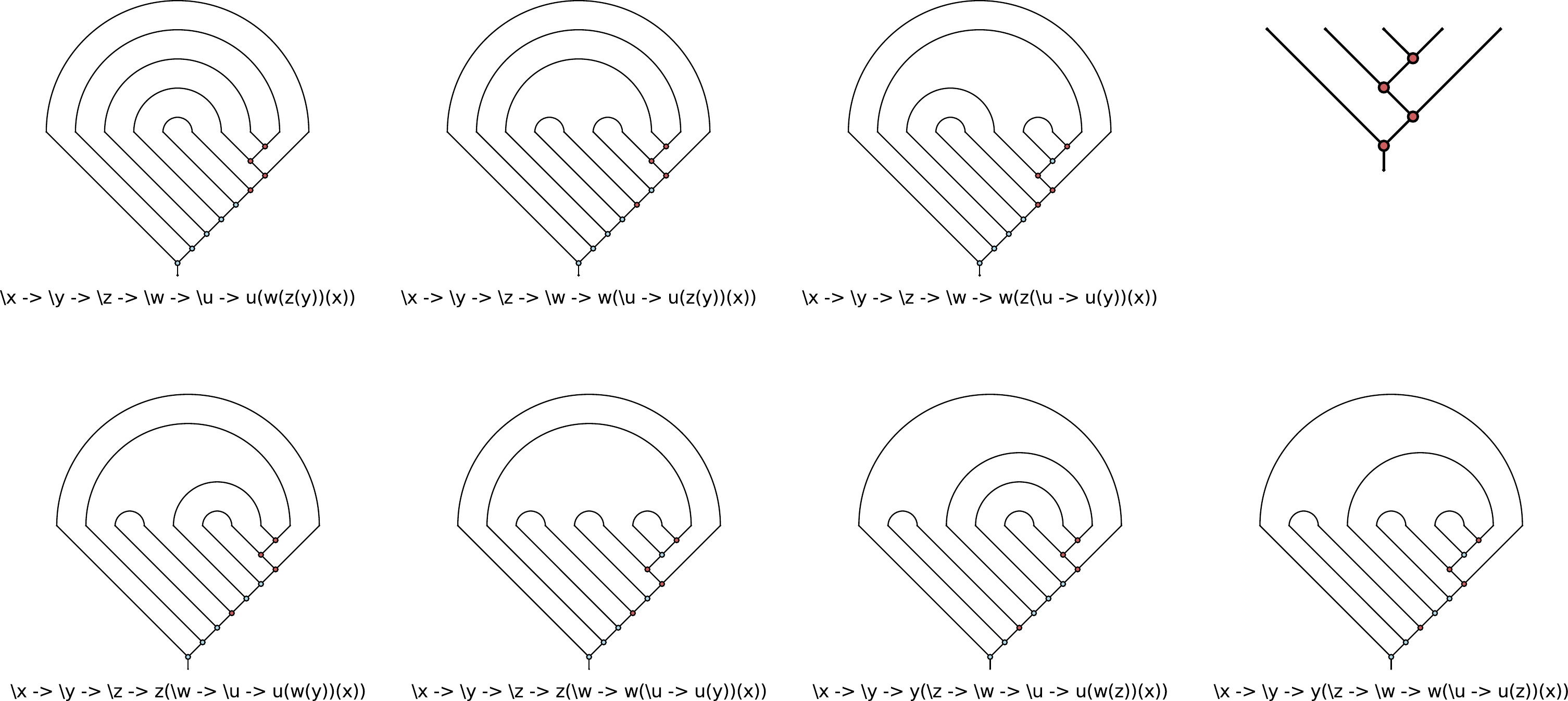}}
\caption{A list of indecomposable normal planar terms of size four, all sharing the same application tree.}
\label{fig:lam4-poi}
\end{center}
\end{figure}

\end{document}